\documentclass[12pt]{article}
\usepackage{geometry}                % See geometry.pdf to learn the layout options. There are lots.close 
\usepackage{graphicx}
\usepackage{amsmath}
\usepackage{xspace} % To avoid problems with missing or double spaces after
\usepackage{amssymb}
\usepackage{epsfig}
\usepackage{epstopdf}
\usepackage{subfigure}
\usepackage{ifthen} % for conditional statements
\newboolean{pdflatex}
\newboolean{inbibliography}
\setboolean{inbibliography}{false} %True once you enter the bibliography
\newboolean{articletitles}
\setboolean{articletitles}{true} % False removes titles in references
\newboolean{uprightparticles}
\setboolean{uprightparticles}{false} %True for upright particle symbols
\usepackage{xspace} 
\usepackage{upgreek}

\newcommand{\offsetoverline}[2][0.1em]{\kern #1\overline{\kern -#1 #2}}%

\def\lhcb   {\mbox{LHCb}\xspace}

\def\cern {\mbox{CERN}\xspace}
\def\lhc    {\mbox{LHC}\xspace}

\def\velo   {VELO\xspace}
\def\rich   {RICH\xspace}

\def\MagUp {\mbox{\em Mag\kern -0.05em Up}\xspace}

\def\hltone {HLT1\xspace}
\def\hlttwo {HLT2\xspace}

\ifthenelse{\boolean{uprightparticles}}%
{

 \def\Ppi         {\ensuremath{\uppi}\xspace}

 \def\Ppsi        {\ensuremath{\uppsi}\xspace}

 \def\PDelta      {\ensuremath{\Delta}\xspace}                 
 \def\PXi         {\ensuremath{\Xi}\xspace}                 
 \def\PLambda     {\ensuremath{\Lambda}\xspace}                 
 \def\PSigma      {\ensuremath{\Sigma}\xspace}                 
 \def\POmega      {\ensuremath{\Omega}\xspace}                 
 \def\PUpsilon    {\ensuremath{\Upsilon}\xspace}

 \def\PB      {\ensuremath{\mathrm{B}}\xspace}                 
                  
 \def\PD      {\ensuremath{\mathrm{D}}\xspace}

 \def\PJ      {\ensuremath{\mathrm{J}}\xspace}                 
 \def\PK      {\ensuremath{\mathrm{K}}\xspace}

 \def\Pc      {\ensuremath{\mathrm{c}}\xspace}

 \def\Pi      {\ensuremath{\mathrm{i}}\xspace}

}
{

 \def\Ppi         {\ensuremath{\pi}\xspace}

 \def\Ppsi        {\ensuremath{\psi}\xspace}                 
                  
 \mathchardef\PDelta="7101
 \mathchardef\PXi="7104
 \mathchardef\PLambda="7103
 \mathchardef\PSigma="7106
 \mathchardef\POmega="710A
 \mathchardef\PUpsilon="7107
                  
 \def\PB      {\ensuremath{B}\xspace}                 
                  
 \def\PD      {\ensuremath{D}\xspace}

 \def\PJ      {\ensuremath{J}\xspace}                 
 \def\PK      {\ensuremath{K}\xspace}

 \def\Pc      {\ensuremath{c}\xspace}

 \def\Pi      {\ensuremath{i}\xspace}

}

\DeclareRobustCommand{\optbar}[1]{\shortstack{{\miniscule (\rule[.5ex]{1.25em}{.18mm})}
  \\ [-.7ex] $#1$}}

\def\cquark    {{\ensuremath{\Pc}}\xspace}
\def\cquarkbar {{\ensuremath{\overline \cquark}}\xspace}
\def\ccbar     {{\ensuremath{\cquark\cquarkbar}}\xspace}

\def\pion   {{\ensuremath{\Ppi}}\xspace}

\def\pip    {{\ensuremath{\pion^+}}\xspace}
\def\pim    {{\ensuremath{\pion^-}}\xspace}

\def\kaon    {{\ensuremath{\PK}}\xspace}
  \def\Kbar    {{\kern 0.2em\overline{\kern -0.2em \PK}{}}\xspace}

\def\KorKbar {\kern 0.18em\optbar{\kern -0.18em K}{}\xspace}

\def\KS      {{\ensuremath{\kaon^0_{\mathrm{S}}}}\xspace}

  \def\Dbar    {{\kern 0.2em\overline{\kern -0.2em \PD}{}}\xspace}
\def\D       {{\ensuremath{\PD}}\xspace}

\def\DorDbar {\kern 0.18em\optbar{\kern -0.18em D}{}\xspace}
\def\Dz      {{\ensuremath{\D^0}}\xspace}

\def\Bbar    {{\ensuremath{\kern 0.18em\overline{\kern -0.18em \PB}{}}}\xspace}

\def\BorBbar    {\kern 0.18em\optbar{\kern -0.18em B}{}\xspace}

\def\jpsi     {{\ensuremath{{\PJ\mskip -3mu/\mskip -2mu\Ppsi\mskip 2mu}}}\xspace}

\def\Y#1S{\ensuremath{\PUpsilon{(#1S)}}\xspace}

\def\Lz          {{\ensuremath{\PLambda}}\xspace}

\def\LorLbar     {\kern 0.18em\optbar{\kern -0.18em \PLambda}{}\xspace}

\newcommand{\decay}[2]{\mbox{\ensuremath{#1\!\to #2}}\xspace}         % {\Pa}{\Pb \Pc}

\def\to                 {\ensuremath{\rightarrow}\xspace}

\def\AT#1     {\ensuremath{A_{\mathrm{T}}^{#1}}\xspace}           % 2

\def\C#1      {\ensuremath{\mathcal{C}_{#1}}\xspace}                       % 9
\def\Cp#1     {\ensuremath{\mathcal{C}_{#1}^{'}}\xspace}                    % 7
\def\Ceff#1   {\ensuremath{\mathcal{C}_{#1}^{\mathrm{(eff)}}}\xspace}        % 9  
\def\Cpeff#1  {\ensuremath{\mathcal{C}_{#1}^{'\mathrm{(eff)}}}\xspace}       % 7
\def\Ope#1    {\ensuremath{\mathcal{O}_{#1}}\xspace}                       % 2
\def\Opep#1   {\ensuremath{\mathcal{O}_{#1}^{'}}\xspace}                    % 7

\newcommand{\tev}{\ifthenelse{\boolean{inbibliography}}{\ensuremath{~T\kern -0.05em eV}}{\ensuremath{\mathrm{\,Te\kern -0.1em V}}}\xspace}
\newcommand{\gev}{\ensuremath{\mathrm{\,Ge\kern -0.1em V}}\xspace}
\newcommand{\mev}{\ensuremath{\mathrm{\,Me\kern -0.1em V}}\xspace}
\newcommand{\kev}{\ensuremath{\mathrm{\,ke\kern -0.1em V}}\xspace}
\newcommand{\ev}{\ensuremath{\mathrm{\,e\kern -0.1em V}}\xspace}
\newcommand{\mevc}{\ensuremath{{\mathrm{\,Me\kern -0.1em V\!/}c}}\xspace}
\newcommand{\gevc}{\ensuremath{{\mathrm{\,Ge\kern -0.1em V\!/}c}}\xspace}
\newcommand{\mevcc}{\ensuremath{{\mathrm{\,Me\kern -0.1em V\!/}c^2}}\xspace}
\newcommand{\gevcc}{\ensuremath{{\mathrm{\,Ge\kern -0.1em V\!/}c^2}}\xspace}
\newcommand{\gevgevcc}{\ensuremath{{\mathrm{\,Ge\kern -0.1em V^2\!/}c^2}}\xspace} % for \pt^2 in CEP
\newcommand{\gevgevcccc}{\ensuremath{{\mathrm{\,Ge\kern -0.1em V^2\!/}c^4}}\xspace} % for q^2

\def\m    {\ensuremath{\mathrm{ \,m}}\xspace}

\def\cm   {\ensuremath{\mathrm{ \,cm}}\xspace}

\def\mm   {\ensuremath{\mathrm{ \,mm}}\xspace}

\def\mum  {\ensuremath{{\,\upmu\mathrm{m}}}\xspace}

\def\nm   {\ensuremath{\mathrm{ \,nm}}\xspace}

\def\gsim{{~\raise.15em\hbox{$>$}\kern-.85em
          \lower.35em\hbox{$\sim$}~}\xspace}
\def\lsim{{~\raise.15em\hbox{$<$}\kern-.85em
          \lower.35em\hbox{$\sim$}~}\xspace}

\def\sqsnn {\ensuremath{\protect\sqrt{s_{\scriptscriptstyle\rm NN}}}\xspace}

\def\mrad{\ensuremath{\mathrm{ \,mrad}}\xspace}

\newcommand{\lum} {\ensuremath{\mathcal{L}}\xspace}
\def\tell1  {TELL1\xspace}
\def\ukl1   {UKL1\xspace}

\newcommand{\ie}{\mbox{\itshape i.e.}\xspace}

\usepackage{booktabs}

\def\pp   {\ensuremath{pp}\xspace}
\def\pH  {\ensuremath{p\text{H}}\xspace}\def\pHe  {\ensuremath{p\text{He}}\xspace}
\def\pNe  {\ensuremath{p\text{Ne}}\xspace}
\def\pAr  {\ensuremath{p\text{Ar}}\xspace}
\def\pppHe{\ensuremath{pp + p\text{He}}\xspace}
\def\pppAr{\ensuremath{pp + p\text{Ar}}\xspace}
\def\papercopyright{\the\year\ CERN for the benefit of the LHCb collaboration} % new since 9/Apr/2018
\def\paperlicence{CC BY 4.0 licence}
\def\paperlicenceurl{https://creativecommons.org/licenses/by/4.0/}

\newcommand{\lhcborcid}[1]{\href{https://orcid.org/#1}{\hspace*{0.1em}\raisebox{-0.45ex}{\includegraphics[width=1em]{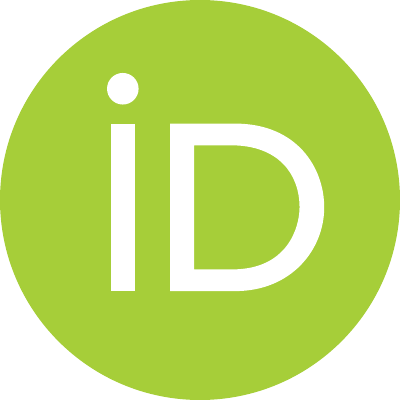}}}}

\usepackage{cite} % Allows for ranges in citations
\usepackage{mciteplus}

\usepackage{rotating}
\usepackage{color}

\usepackage{hyperref}
\usepackage{hyperxmp}
\usepackage[all]{hypcap} % Internal hyperlinks to floats.
\usepackage{comment}
\usepackage{lineno}

\begin{document}

% Header ---------------------------------------------------
\vspace*{-1.5cm}
\centerline{\large EUROPEAN ORGANIZATION FOR NUCLEAR RESEARCH (CERN)}
\vspace*{1.5cm}
\noindent
\begin{tabular*}{\linewidth}{lc@{\extracolsep{\fill}}r@{\extracolsep{0pt}}}
\ifthenelse{\boolean{pdflatex}}% Logo format choice
{\vspace*{-1.5cm}\mbox{\!\!\!\includegraphics[width=.14\textwidth]{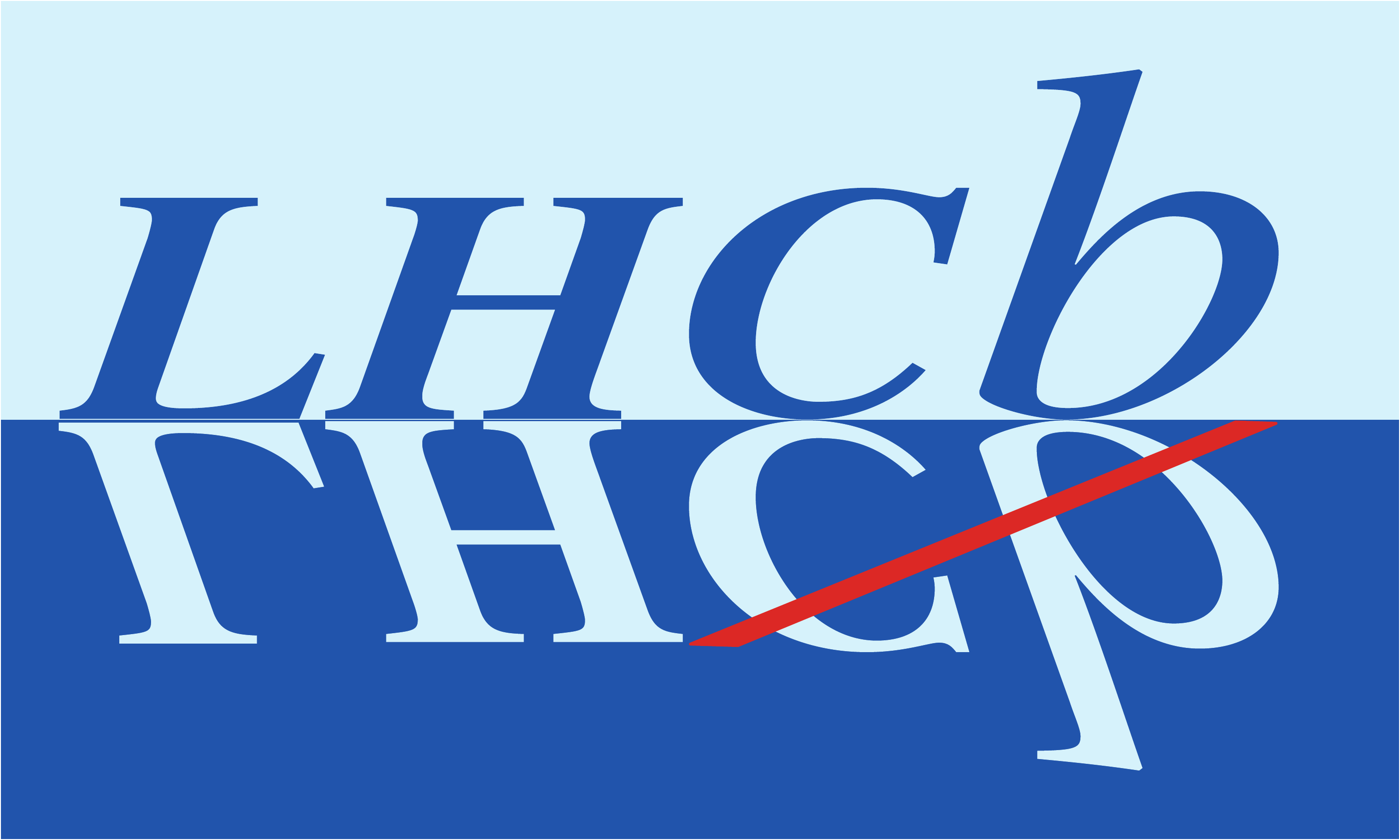}} & &}%
{\vspace*{-1.2cm}\mbox{\!\!\!\includegraphics[width=.12\textwidth]{figs/lhcb-logo.pdf}} & &}%
\\
 & & LHCb-DP-2024-002 \\  % ID 
 & & November 8, 2024 \\ % Date - Can also hardwire e.g.: 23 March 2010
 & & \\
% not in paper \hline
\end{tabular*}

\vspace{0.5cm}

{\normalfont\bfseries\boldmath\huge
\begin{center}
High-density gas target at the LHCb experiment
\end{center}
}

\vspace{0.5cm}

%\linenumbers

%\begin{center}
\begin{flushleft}

% Authors -------------------------------------------------
O.~Boente~Garcia$^1$\lhcborcid{0000-0003-0261-8085},
G.~Bregliozzi$^2$, 
D.~Calegari$^2$,
V.~Carassiti$^3$\lhcborcid{0000-0002-5471-5533}, 
G.~Ciullo$^{3,4}$\lhcborcid{0000-0001-8297-2206}, 
V.~Coco$^2$\lhcborcid{0000-0002-5310-6808},
P.~Collins$^2$\lhcborcid{0000-0003-1437-4022},
P.~Costa~Pinto$^2$, 
C.~De~Angelis$^{5,6}$\lhcborcid{0009-0005-5033-5866}, 
P.~Di~Nezza$^{7*}$\lhcborcid{0000-0003-4894-6762},
R.~Dumps$^2$, 
M.~Ferro-Luzzi$^2$\lhcborcid{0009-0008-1868-2165},
F.~Fleuret$^1$\lhcborcid{0000-0002-2430-782X},
G.~Graziani$^8$\lhcborcid{0000-0001-8212-846X},
S.~Kotriakhova$^{3,4}$\lhcborcid{0000-0002-1495-0053},
P.~Lenisa$^{3,4}$,\lhcborcid{0000-0003-3509-1240}
Q.~Lu$^1$\lhcborcid{0000-0002-6598-1941},
C.~Lucarelli$^{8,9}$\lhcborcid{0000-0002-8196-1828},
E.~Maurice$^1$\lhcborcid{0000-0002-7366-4364},
S.~Mariani$^{2*}$\lhcborcid{0000-0002-7298-3101},
K.~Mattioli$^1$\lhcborcid{0000-0003-2222-7727},
M.~Milovanovic$^2$\lhcborcid{0000-0003-1580-0898},
L.L.~Pappalardo$^{3,4}$\lhcborcid{0000-0002-0876-3163},
D.M.~Parragh$^2$\lhcborcid{0000-0003-4963-4381},
A.~Piccoli$^{3,4}$\lhcborcid{0009-0008-3313-7413},
P.~Sainvitu$^2$,
B.~Salvant$^2$\lhcborcid{0000-0003-1281-5509},
F.~Sanders$^{12}$,
M.~Santimaria$^7$\lhcborcid{0000-0002-8776-6759},
J.~Sestak$^2$,
S.~Squerzanti$^3$,
E.~Steffens$^{10}$\lhcborcid{0000-0003-4029-3154},
G.~Tagliente$^{11}$\lhcborcid{0000-0003-1665-4191},
W.~Vollenberg$^2$,
C.~Vollinger$^2$\lhcborcid{0009-0000-4993-0987}

\bigskip

{\normalfont\itshape\footnotesize
$ ^1$ Laboratoire Leprince-Ringuet, CNRS/IN2P3, Ecole Polytechnique, Institut Polytechnique de Paris, Palaiseau, France \\
$ ^2$ European Organization for Nuclear Research (CERN), Geneva, Switzerland \\
$ ^3$ INFN Sezione di Ferrara, Ferrara, Italy \\
$ ^4$ University of Ferrara, Ferrara, Italy \\
$ ^5$ University of Cagliari, Cagliari, Italy \\
$ ^6$ INFN Sezione di Cagliari, Cagliari, Italy \\
$ ^7$ INFN Laboratori Nazionali di Frascati, Frascati, Italy \\
$ ^8$ INFN Sezione di Firenze, Firenze, Italy \\
$ ^{9}$ University of Firenze, Firenze, Italy \\
$ ^{10}$ Friedrich-Alexander-Universit{\"a}t Erlangen-N{\"u}rnberg (FAU), Erlangen-N{\"u}rnberg, Germany \\
$ ^{11}$ INFN Sezione di Bari, Bari, Italy \\
$ ^{12}$ Nikhef National Institute for Subatomic Physics, Amsterdam, Netherlands \\
}
%\end{center}
\medskip
\footnotesize{$^*$~contact~authors: dinezza@infn.it, saverio.mariani@cern.ch}

\end{flushleft}

% Abstract -----------------------------------------------
\begin{abstract}
\noindent
The recently installed internal gas target at \lhcb presents exceptional opportunities for an extensive physics program for heavy-ion, hadron, spin, and astroparticle physics. A storage cell placed in the LHC primary vacuum, an advanced Gas Feed System, the availability of multi-TeV proton and ion beams and the recent upgrade of the \lhcb detector make this project unique worldwide.
In this paper, we outline the main components of the system, the physics prospects it offers and the hardware challenges encountered during its implementation. The commissioning phase has yielded promising results, demonstrating that fixed-target collisions can occur concurrently with the collider mode without compromising efficient data acquisition and high-quality reconstruction of beam-gas and beam-beam interactions.
\end{abstract}
\begin{center}
Published on \href{https://link.aps.org/doi/10.1103/PhysRevAccelBeams.27.111001}{Physical Review Accelerators and Beams 27,
111001 (2024)} 

DOI: 10.1103/PhysRevAccelBeams.27.111001
\end{center}

\vspace{\fill}

{\footnotesize 
% Edit macro in main.tex to keep metadata correct
\centerline{\copyright~\papercopyright. \href{\paperlicenceurl}{\paperlicence}.}}
\vspace*{2mm}

\cleardoublepage
\tableofcontents
\cleardoublepage
\section{Introduction}
%\section{Introduction}
%\section{Introduction} \label{sec:introduction}
\label{sec:introduction}

\lhcb is the first experiment at the \lhc that can operate simultaneously with two distinct interaction regions. As part of the major \lhcb Upgrade~\cite{LHCb-DP-2022-002}, SMOG2~\cite{LHCb-TDR-020}, the new fixed‑target system replacing the previous System for Measuring Overlap with Gas (SMOG)~\cite{Thesis_SMOG}, was installed during the \lhc long shutdown 2 (LS2). The core of the system is a gas target concentrated within a 20 cm‑long aluminum storage cell. It is positioned at the upstream edge of the \lhcb vertex locator  (\velo)~\cite{LHCb-TDR-013, PCollins_VELO}, the silicon pixel tracker closest to the beam, 34 cm away from the main interaction point, and coaxial with the \lhc beamline. 

The storage cell technology~\cite{Erhard_Steffens_2003} allows for controlled injection of a limited amount of gas into a well-defined volume within the \lhc beam pipe. This control ensures precise management of the gas pressure and density distribution while maintaining the vacuum level of the beam pipe at least two orders of magnitude below the upper limit required for \lhc operations.
With beam‑gas interactions occurring at approximately 4\% of the proton–proton (\pp) collision rate at \lhcb, the beam’s lifetime remains largely unaffected. 

The narrowness of the cell allows for data collection of 100 pb$^{-1}$ of proton fixed-target data per year with a flow rate as low as 10$^{15}$ particles (atoms or molecules, depending on the exploited gas) per second. The new injection system is able to switch between different gases within a few minutes, 
enabling injection of any type of gas compatible with \lhc operation. To date helium (He), neon (Ne), argon (Ar), and hydrogen (H$_2$) have been injected into the beam pipe, but injections of deuterium (D$_2$), nitrogen (N$_2$), oxygen (O$_2$), and possibly heavier noble gases like krypton (Kr) and xenon (Xe) are being studied. With all of these, SMOG2 opens new windows for QCD studies and production measurements  relevant for astroparticle physics at the \lhc, accessing kinematic regions poorly probed before. Combined with \lhcb's excellent particle identification capabilities, and momentum and impact parameter resolutions, the new gas target system will advance the understanding of the gluon, antiquark, and heavy‑quark constituents of nucleons and nuclei at large Bjorken-$x$. The gas target also offers the opportunity to investigate the dynamics and spin distributions of quarks and gluons inside unpolarized nucleons, which has not yet been explored at \lhc. The study of particles produced in collisions with light nuclei, such as H$_2$ and He, and possibly N$_2$ and O$_2$, will provide valuable reference data for cosmic‑ray physics and investigations related to dark matter. Moreover, SMOG2 will enable \lhcb to perform studies of heavy-ion collisions at large rapidities, in an unexplored energy range between the SPS and RHIC, offering new insights into the QCD phase diagram.

This article will discuss the envisaged physics program and the hardware and software solutions adopted for the SMOG2 system, addressing also the challenges associated with the interplay between the storage cell and the stringent requirements imposed by \lhc and \lhcb. The last section presents the data acquired  during commissioning runs performed in 2022, demonstrating the remarkable capabilities of this system.
\section{Physics perspectives}
%\section{Physics}
\label{sec:physics}

Unlike most major colliders in the past, no dedicated fixed-target 
experiments using an extracted beam were foreseen within the \lhc accelerator complex.
However, the \lhc unprecedented energy offers unique opportunities 
for hadronic physics measurements also in fixed-target collisions. 
With a beam energy ranging from 450 to 6800\gev per nucleon, collisions of protons or lead ions on gas targets of different atomic numbers can be obtained 
with a nucleon-nucleon center-of-mass energy \sqsnn\xspace between 29 and 113\gev. Such an energy range, combined with the forward kinematics accessible in the fixed-target configuration, fills a gap between previous fixed-target experiments for example operating at the SPS accelerator ($\sqsnn\sim 10 - 30\gev$) and the heavy-ion collider data by RHIC experiments ($\sqsnn\sim 200 \gev$). Novel physics opportunities, complementary to the exploitation of beam-beam collisions, notably include:

\begin{itemize}
\item access to nucleon and nuclear Parton Distribution Functions (PDFs) at large Bjorken-$x$,
including the charm and beauty quark PDFs;
\item study of nuclear-matter effects, using hydrogen as a reference system to compare with larger nuclear targets such as argon, 
krypton, and xenon; 
\item production studies relevant for cosmic-ray physics. 
Collisions on hydrogen and helium targets reproduce indeed
primary cosmic-ray collisions in the interstellar medium, while data with 
nitrogen or oxygen targets (or proxies like neon) can contribute to the modeling
of extensive showers from Ultra-High-Energy (UHE) cosmic rays in the atmosphere.
\end{itemize}
 
The potential of \lhcb's fixed-target configuration has been already 
demonstrated during the \lhc Run 2 using the SMOG target. The first results include measurements of charm production with \mbox{$\sqsnn=86.6\gev$} \pHe, \mbox{$\sqsnn=110.4\gev$} \pAr~\cite{LHCb-PAPER-2018-023} and \mbox{$\sqsnn=68.5\gev$} \pNe and PbNe data~\cite{LHCb-PAPER-2022-015, LHCb-PAPER-2022-014,LHCb-PAPER-2022-011}, and measurements of antiproton production in \mbox{$\sqsnn=110.4\gev$} \pHe collisions~\cite{LHCb-PAPER-2018-031,LHCb-PAPER-2022-006}.

SMOG2 provides a breakthrough in the achievable integrated luminosity for all these studies. 
The combined effect of the increased gas target density and the possibility to acquire fixed-target data routinely, concurrently with the
standard \lhcb data-taking with beam-beam collisions, will result in recorded samples
of  beam-gas collisions corresponding to integrated luminosities of order 100~pb$^{-1}$ per year.
This is comparable to the dedicated high-energy fixed-target experiments performed at previous accelerator facilities, like the Tevatron or the SPS. Also, owing to the direct luminosity measurement and to the  confinement of the gas upstream of the \velo detector, increased experimental efficiency with respect to the \lhc Run~2 is expected. Finally, by varying the injected gas from hydrogen to krypton or xenon, denser collision systems, and hence richer hadronic environments, can be explored.  

The resulting physics opportunities, discussed in more detail in~\cite{LHCb-PUB-2018-015}, are summarized in the following.

\subsection{Heavy-ion physics}
Relativistic heavy-ion collisions give access to the high-temperature and high-density regime of QCD, where the production of heavy-quarks is well suited to study the transition between ordinary hadronic matter and the hot and dense Quark-Gluon Plasma (QGP), the regime where partons are asymptotically free from color confinement. Since heavy-quark masses are large compared to the QGP critical temperature $T_c\sim 156\mev$ \cite{PLB795}, their production occurs in primary nucleon-nucleon collisions, at an early stage of the interaction. They can therefore experience the full evolution of the created nuclear medium, including the deconfinement phase. The latter is expected to significantly affect the formation of hidden heavy-flavor bound states with respect to the overall heavy-flavor production \cite{PRD50}. This so-called color screening effect is considered one of the key signatures of QGP formation. As the screening temperature depends on the radius of the quark-antiquark bound state, a larger suppression is expected for excited states, an effect known as sequential suppression \cite{PLB637}. The \lhcb detector gives the opportunity to measure, for the first time, hidden and open charm hadrons, including 1P states, in heavy-ion collisions where the contribution from charm quark recombination is expected to be negligible. 
While indeed at the highest energies the contribution from recombination of charm-quark pairs during the deconfined QGP phase needs to be taken into account, it is expected to be negligible in fixed-target configuration, as on average one \ccbar pair only is produced. This gives the \lhcb heavy-ion fixed-target program the unique opportunity to probe and explore the full sequential suppression pattern of charmonia, so far unobserved experimentally.

\subsection{Nucleon structure}
The nucleon structure is traditionally parametrized in terms of PDFs, which, in their simplest (collinear) form, are functions of the longitudinal momentum fraction of quarks and gluons, expressed by the Bjorken-$x$ variable. Although tremendous advances have been made over the past decades in defining the quark and gluon dynamical substructure of the nucleon, the present knowledge of the PDFs still suffers from large uncertainties, especially at very-high and very-low $x$~\cite{nnpdf}, leaving open fundamental questions about QCD and confinement. In many cases, the PDF uncertainties have become the limiting factor in the accuracy of the predictions for \lhc measurements~\cite{wz,Delgado}.

Our understanding of collinear PDFs has primarily been derived from inclusive deep inelastic scattering (DIS) experiments~\cite{Armesto_2024}. However, by considering the explicit dependence of PDFs on parton transverse momenta, a new perspective has emerged in exploring the nucleon structure (for reviews, see Refs.~\cite{TMD-Rev1, TMD-Rev2}). These transverse-momentum-dependent PDFs (TMDs) have opened up avenues to study spin-orbit correlations within the nucleon and provide insights into the elusive parton orbital angular momentum, a critical piece in understanding the proton spin puzzle~\cite{TMD-Rev1}. Additionally, TMDs offer the opportunity to map parton densities in three dimensions, akin to nucleon tomography in momentum space.

There are two quark TMDs involved in unpolarized processes: the standard unpolarized distribution function $f_1^q$ and the poorly known Boer-Mulders function $h_1^{\perp, q}$~\cite{BM}. Even if it requires no beam or target polarization, the Boer-Mulders function is in fact a polarized TMD because it describes the correlation between the quark transverse polarization and transverse momentum. In the last 20 years, significant progress has been achieved in the comprehension of the quarks TMDs in Semi-Inclusive DIS (SIDIS) experiments (HERMES, COMPASS, JLab)~\cite{TMD-Rev3}. Proton collisions at \lhc constitute a complementary approach as they can access significantly higher energy scales than any data from existing fixed-target experiment. Furthermore, by comparing the results obtained in SIDIS and in hadronic collisions, it is possible to perform fundamental tests of QCD factorization, evolution, and universality. 

In contrast to the quark TMDs, the present knowledge of the gluon TMDs is very poor. Although the theoretical framework is well consolidated, the experimental access is still extremely limited. Similarly to the quark case, two gluon TMDs appear in unpolarized observables: the spin-independent function $f_1^g$ and the linearly-polarized gluon TMD $h_1^{\perp, g}$. The latter is particularly interesting since, in analogy to the Boer-Mulders function, it carries information on the gluon (linear) polarization in an unpolarized proton. Both distribution functions are process dependent~\cite{Boer16} and can test QCD universality once compared with the analogous measurements in $ep$ collisions, e.g. at a future Electron-Ion Collider~\cite{EIC_1, EIC_2}.

The quark $f_1^q$ and $h_1^{\perp, q}$ TMDs can be probed in Drell-Yan processes, exploiting the excellent reconstruction capabilities of \lhcb for muon-pairs. Furthermore, by feeding the SMOG2 system with either H$_2$  and D$_2$, sensitivity to both the $u$ and $d$ quark contributions is obtained. Another important reason to study unpolarized Drell-Yan processes is to get access to the antiquark content of the nucleon. More specifically, by using H$_2$ and D$_2$ targets with SMOG2, the poorly constrained antiquark momentum distributions $\bar{u}(x)$ and $\bar{d}(x)$ can both be accessed, complementing the recently published surprising E906 results~\cite{E906}. Last but not least is the study of the Generalized Parton Distribution functions by measuring Ultra Peripheral Collision events~\cite{LHCb-PAPER-2022-012}.

\subsection{Measurements relevant to cosmic-ray physics } 

In recent years, space-based cosmic-ray detectors~\cite{pamelaPbar, amsPbar} have dramatically improved our knowledge of the cosmic-ray composition for energies up to 500\gev. Measurements of cosmic antimatter constitute an indirect probe for dark matter annihilation or other exotic antimatter sources. These searches are presently limited in accuracy by the 
knowledge of the cross-section of antimatter production in collisions of cosmic rays with the interstellar medium~\cite{Donato}. This is essentially composed of hydrogen ($\sim$90\%) and helium ($\sim$10\%), therefore the SMOG2 configuration with H$_2$ or He target is ideal to reproduce these collisions at the needed energy scale. 
Using also a D$_2$ target, differences in antiproton production between $pp$ and $pn$ collisions can be precisely quantified,
constraining the difference between antiproton and antineutron production. 
Systematic uncertainties are expected to improve significantly with respect to past measurements with SMOG due to the better determination of the luminosity achievable with the precise calibration of the SMOG2 gas injection.
It is also planned to take data at different beam energies to study the energy evolution of the antiproton production cross-section, to precisely constrain the violation of
Feynman scaling and study the enhancement of antihyperon production. Data at lower beam energy will also provide a wider coverage towards forward rapidities in the center-of-mass frame.

Measurements with atmosphere-like targets (N$_2$, O$_2$, Ne) can contribute to the understanding of UHE 
cosmic showers in the atmosphere. While \lhc data in collision mode provides access to an energy
scale corresponding to the first collision of $10^{17}$~eV cosmic rays, data over many orders of magnitudes
are needed to model the full shower development. 
Data produced by SMOG2 are expected to contribute to 
the interpretation of the muon lateral profile measurements in UHE showers, where data diverge significantly from 
model predictions~\cite{muonPuzzle}.  
While the \lhcb acceptance covers central and backward center-of-mass rapidities ($-3 \lesssim y^*  \lesssim 0$)  
for proton-on-oxygen data,
the planned \lhc run with oxygen beams, using hydrogen as a target, can be exploited to access the forward
rapidities~\cite{oxygen}.

\section{The storage cell}
%\section{Storage Cell Gas Target}
\label{sec:sc}

\subsection{Principle of operation}
\label{sec:sc_principle}

The storage cell technique, originally proposed by W. Haeberli (1925 - 2021) in 1965, was successfully demonstrated by his group at the University of Wisconsin in 1980~\cite{SC_1}. Since then, this method has been applied, mainly for polarized targets, in several experiments~\cite{SC_2, SC_3, SC_4, PhysRevSTAB.18.020101, PhysRevSTAB.9.050101}.

\begin{figure}
    \centering
    \includegraphics [width=0.7\linewidth] {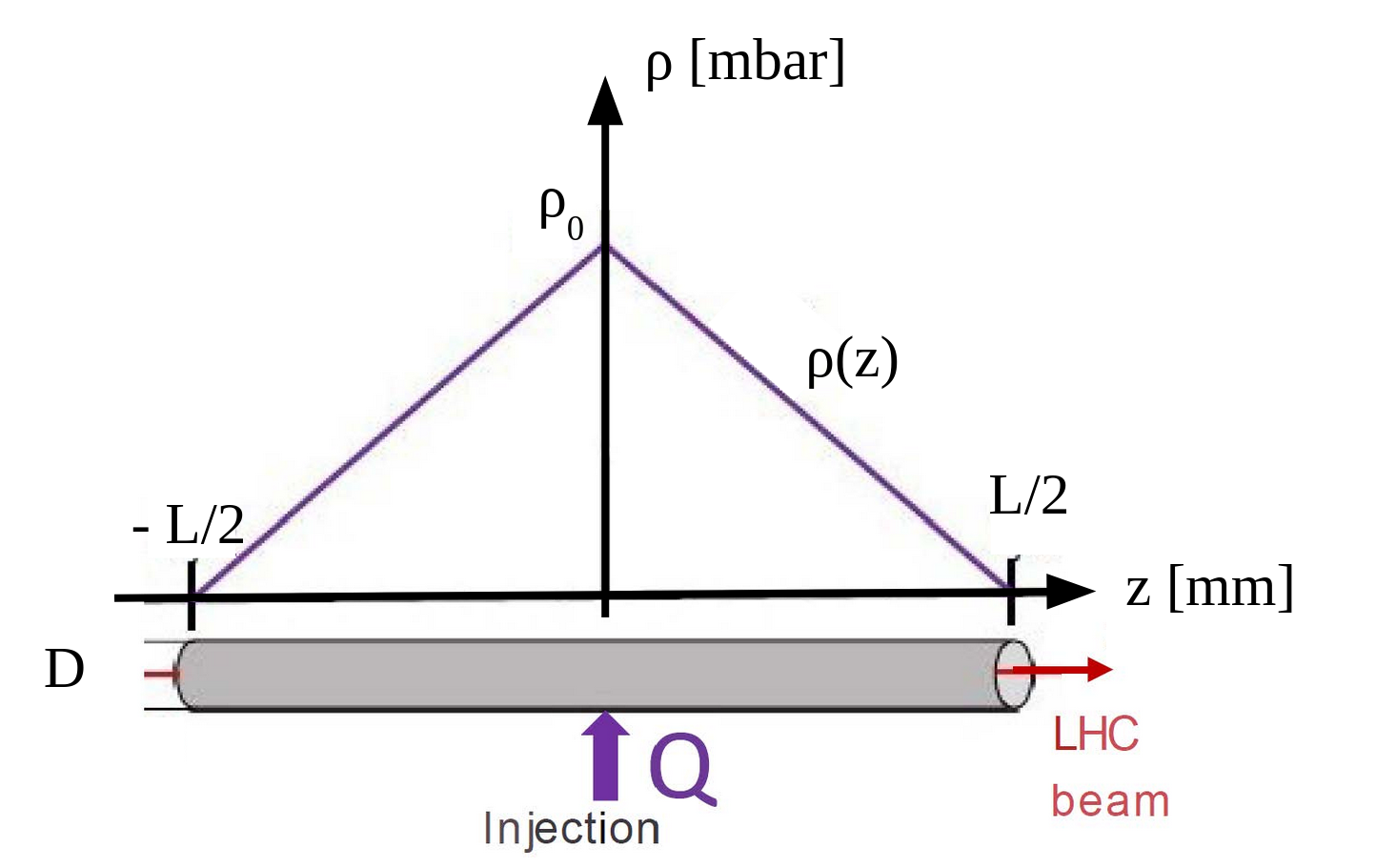}
    \caption{Scheme of a tubular storage cell of length L and inner diameter D. Injection occurs in the center with flow rate Q, resulting in a triangular density distribution $\rho(z)$ with maximum $\rho_0$ at the center.}
    \label{fig:sc}
\end{figure}

The storage cell consists of an open-ended tube positioned around the beam path, as schematically shown in Fig.~\ref{fig:sc}. Gas is injected at the center of the tube from where the molecules or atoms diffuse towards both ends. This process allows to obtain a density up to 2 orders of magnitude higher with respect to direct injections into the \velo beam vacuum, for the same flow rate, and over a shorter length along the beamline.

In a storage cell, the gas produces a triangular pressure profile with maximal density $\rho_0$ at
the center and a target areal density $\theta = \rho_0 \cdot L/2$.
At the typical densities used in the SMOG2 storage cell, gas diffusion occurs in the molecular flow regime, where wall collisions dominate and re-emissions angles follow the Knudsen's cosine law~\cite{Knudsen}.
The flow rate and the corresponding volume density can be determined through (i) the Analytic Method  (AM) employing parameters such as geometry, molecular mass, and wall temperature as parameters, or with (ii) Numerical Simulations (Simu), such as the \textit{Molflow+} program~\cite{MolFlow} (see Sec.~\ref{sec:Molflow}).

\subsection{Mechanical design and construction}
\begin{figure}
    \centering
    \includegraphics [width = \linewidth] {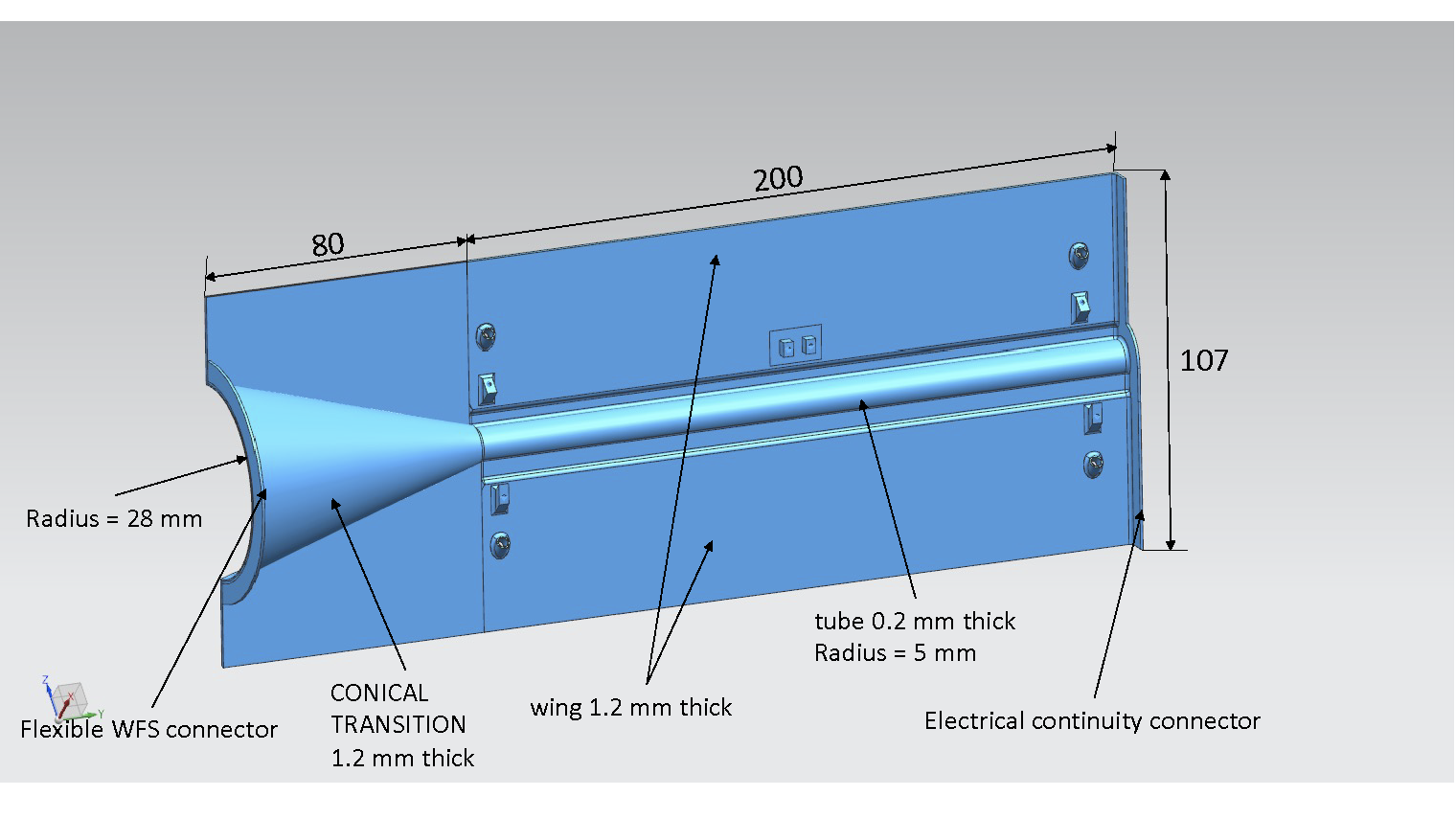}
    \caption{Dimensions of one half of the cell and its transition cone pointing to the upstream side of the \velo.}
    \label{fig:halfcell}
\end{figure}

Mechanically, the SMOG2 cell consists of two halves, rigidly connected to the two \velo detector boxes. Due to the large transverse size of the \lhc beam at the injection energy of 450 GeV, the cell is kept open together with the \velo boxes during beam injection and tuning, and closed once the stable beam condition is reached.
The core of the storage cell consists of a tube connected on one side to the upstream beam pipe and on the other side to the \velo Radio-Frequency (RF) box~\cite{PCollins_VELO}. The tube has a length of 20\cm, an inner diameter of 1\cm (in the closed position), and a wall thickness of 200 $\mu$m. It is followed by a short conical extension, made out of the same piece of aluminum, allowing the diameter to be adapted to the one of the upstream beam pipe. Two 5\cm wide side wings provide a lateral sealing.
In Fig.~\ref{fig:halfcell}, the main dimensions of the half cell are reported. 

\begin{figure*}
    \centering
    \includegraphics [width = 1.0\linewidth] {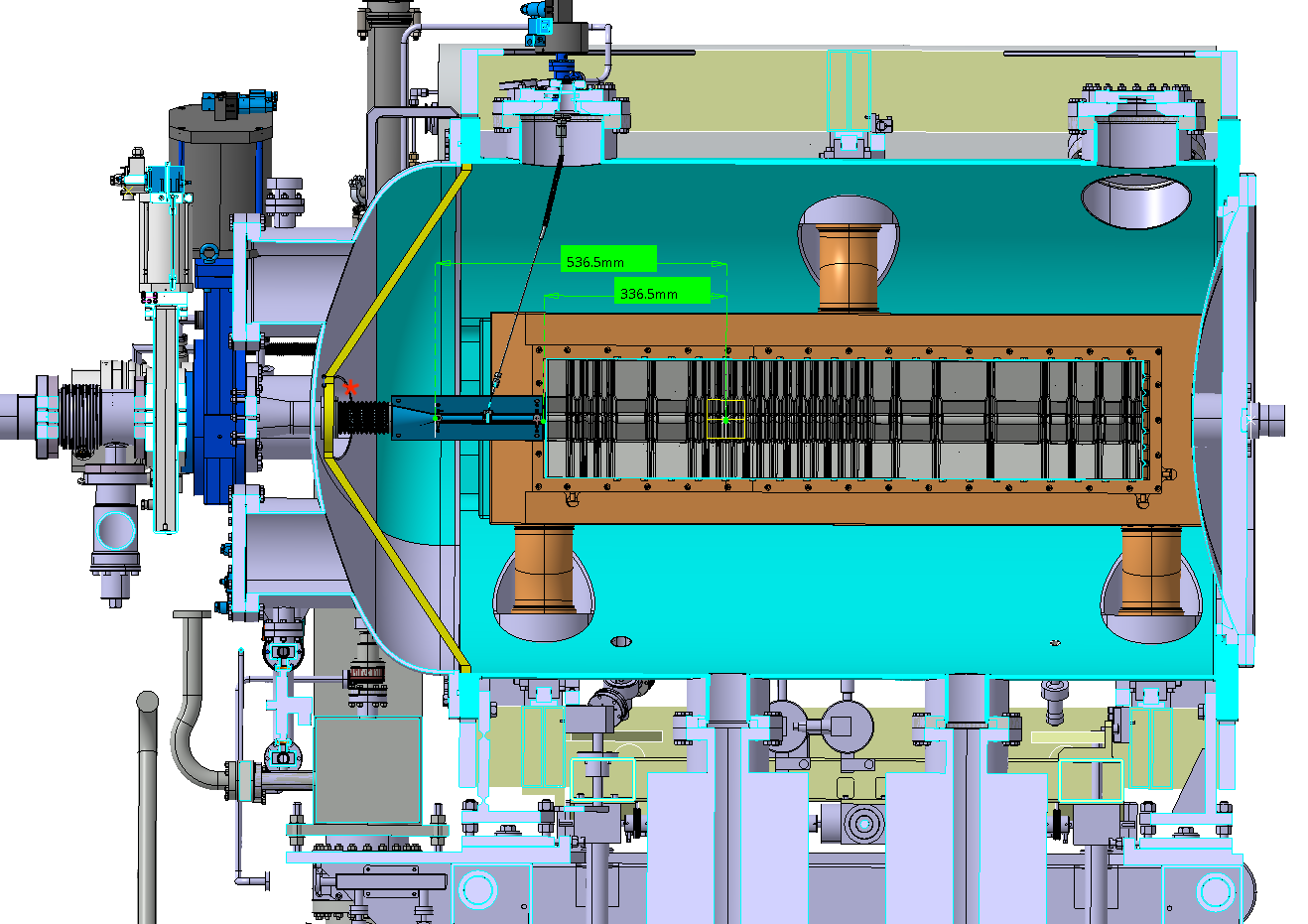}    
    \caption{Overall view of the \velo vessel with the storage cell (in dark blue) positioned just upstream of the RF boxes (light
green). The distances of the cell edges from the beam-beam interaction point are indicated in yellow, covering 200 mm from
-536.5 mm to -336.5.  The connection to the Gas Feed System at the top of the tank can be seen.  The red star indicates the injection position when the injection type is chosen to be as for the previous
SMOG system, as discussed in section V.}
    \label{fig:cell1b}
\end{figure*}

\begin{figure}
    \centering
    \includegraphics [width = \linewidth] {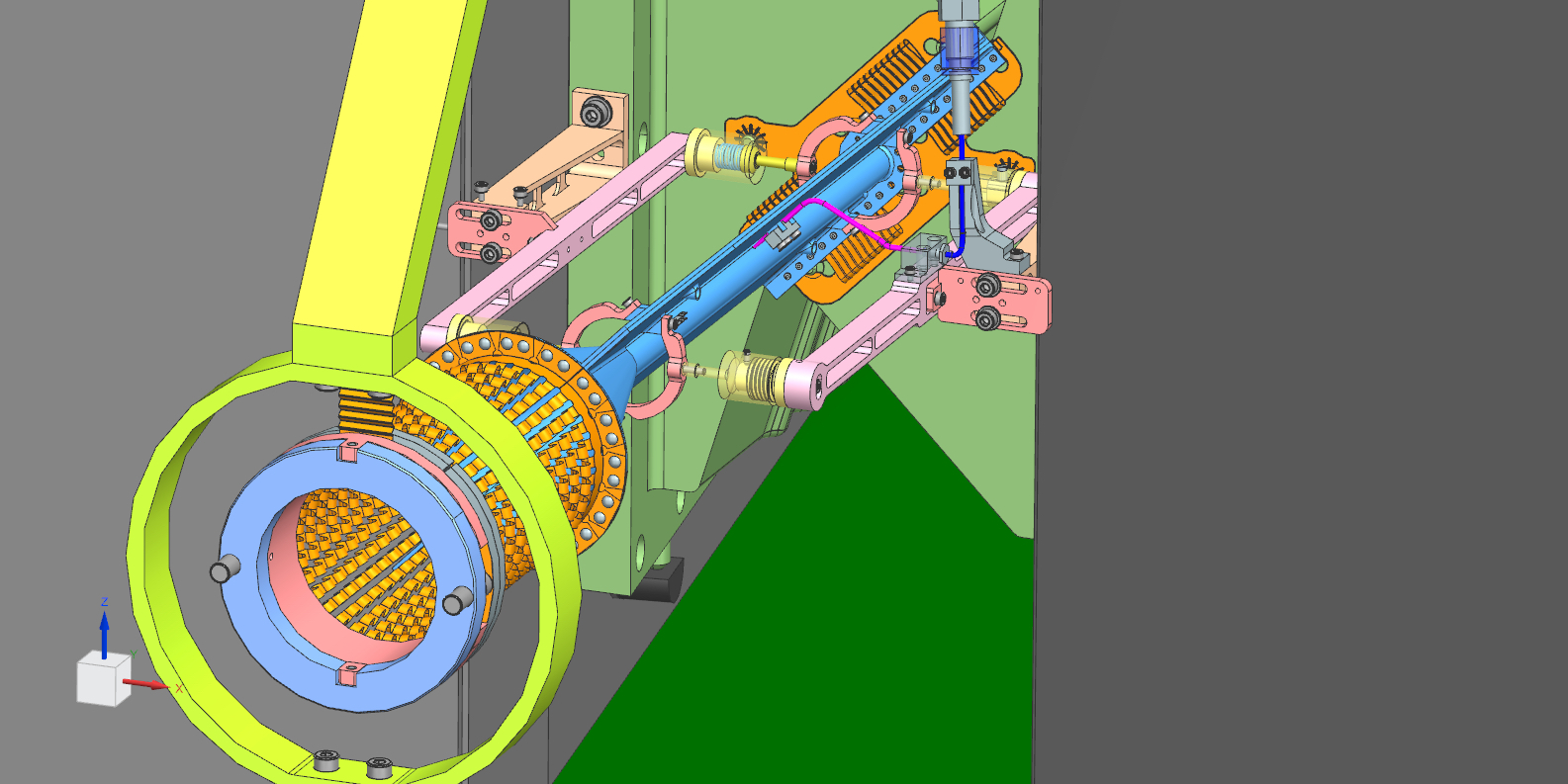}
    \caption{Zoom on the storage cell to show the supports and attachment to the \velo RF boxes and upstream beam pipe ring (light blue) via wake field suppressors (in gold).}
    \label{fig:cell1a}
\end{figure}

Both the cell and its support are made using an aluminum alloy (EN AW-5083: Mg 4~\%, Mn 0.5~\%). Figure~\ref{fig:cell1b} shows the CAD transverse view of the cell installed in the \velo vessel, Fig.~\ref{fig:cell1a} the cell system in its closed position.

To ensure gas containment without any lateral leak and to meet the stringent requirements on the planarity to be within 50\mum, the cell has been realized by milling the shape out of an aluminum block. After completing the outer surface, each half-cell has been accommodated on a vacuum plate to keep its shape flat while the inner surface was finished.
The cell is rigidly mounted to the \velo boxes by two cantilevers screwed to the flange of the \velo RF boxes.
The \velo design foresees the possibility to close the detector with 
a final gap that could deviate from the nominal zero value by 0.1-0.2\mm in order to accommodate possible geometrical imperfections of the complex corrugated faces of the boxes.
To account for this uncertainty, one half of the cell is rigidly fixed to the detector box, 
while the other one is mounted on a spring system that allows for an adequate flexibility when reaching the closed position.
During the installation phase, the alignment system of the fixed half-cell enabled the centering of the cell axis with respect to the \velo detector axis, as described in Sec.~\ref{sec:alignment}. It is worth noting that the \lhc beam always goes through the center of the storage cell, regardless of the beam's position in the machine. In fact, once stable beam conditions are reached, the \velo detector and, consequently, the cell are in place into an optimized position centered around the interaction region in both \textit{x} and \textit{y} coordinates. This position is not known beforehand and it may vary over $\pm 5\mm$ in both \textit{x} and \textit{y}, even from fill to fill. Therefore, a software procedure determines the beam position while the detectors are not yet completely moved in, and then they are adjusted to the optimal position. This is performed with a motion mechanism that can bring the detectors to their \textit{x} and \textit{y} position with an accuracy of the order of 10\mum.

\begin{figure}
    \centering
    \includegraphics [width = \linewidth] {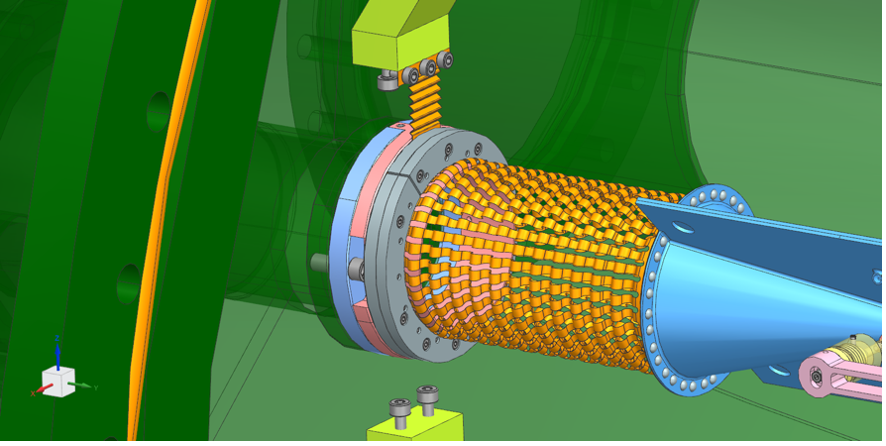} 
    \caption{Details of the upstream WFS and its connection to the beam pipe flange.}
    \label{fig:flange}
\end{figure}

Two Cu-Be2 Wake Field Suppressors (WFS) positioned at the upstream and downstream ends of the cell ensure electrical continuity. 
The use of the Cu-Be2 alloy for the WFS offers a combination of excellent electrical and thermal conductivity. The 0.075\mm thickness provides excellent elasticity to the movement, ensuring mechanical robustness, as well as fatigue resistance.
The upstream WFS has a cylindrical shape, as shown in Fig.~\ref{fig:flange}, and connects the beam tube (56\mm diameter) with the cell tube (10\mm diameter) through the smooth, conical transition, discussed above. The WFS itself consists of two foils cold-formed and wire-eroded strips that act as springs, forming two flexible half-tubes capable of accommodating the motion of the cell. The other edge of the WFS is firmly secured to the cell using aluminum rivets.

The downstream WFS is connected to the cell by small tubular rivets, while the connection of the WFS to the RF box of the \velo detector uses the same technique as the previous WFS.

The gas is injected into the center of the cell through a  1.1\mm inner diameter stainless steel capillary which, on one extremity, is pushed into a 1.47\mm hole in the fixed half-side of the storage cell. A dedicated aluminium support prevents the capillary to protrude the inner surface of the cell. On the other extremity, the capillary is connected to a flexible transition, which extends to the air side via a standard DN16CF flange on the \velo vessel.

A comprehensive fatigue testing program has been conducted on prototypes, subjecting them to over 15000 cycles of repetitive opening and closing. This extensive testing, which corresponds to more than 15 years of operational usage, has revealed no indications of structural alterations. 

\begin{figure*}
	\includegraphics[width=\linewidth]{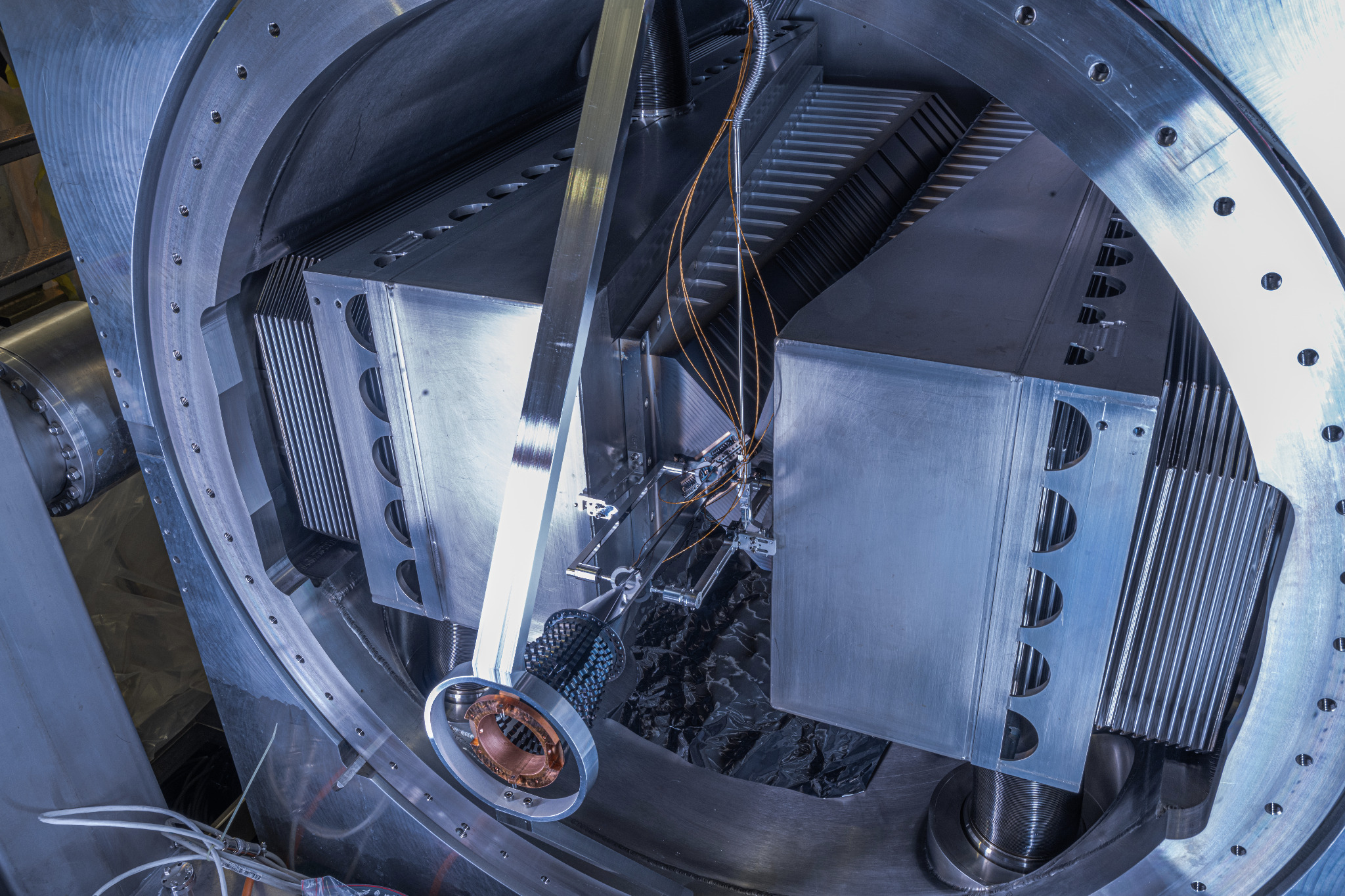}
    \caption{Picture of the storage cell, in closed position, installed in front of the \velo RF foil within the \velo vessel.}
    \label{fig:photo}
\end{figure*}

Figure \ref{fig:photo} shows a picture of the storage cell in front of the \velo RF foil, within the \velo vessel, during the SMOG2 installation.

\subsection{Temperature monitoring system}
\begin{figure}
    \centering
    \includegraphics [width=0.6\linewidth] {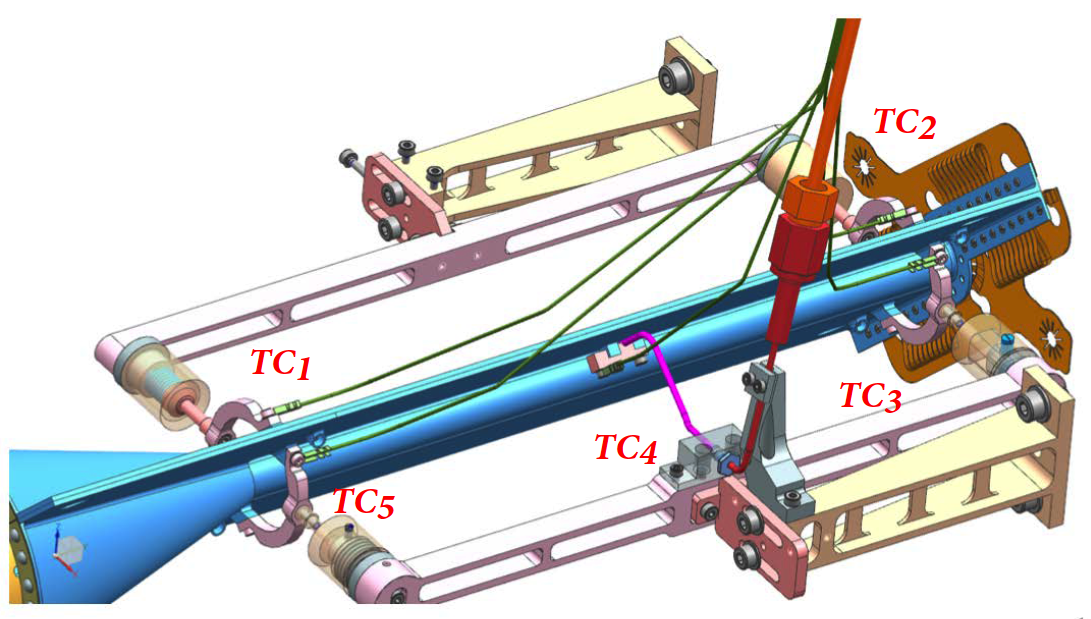} 
    \includegraphics [width=0.6\linewidth] {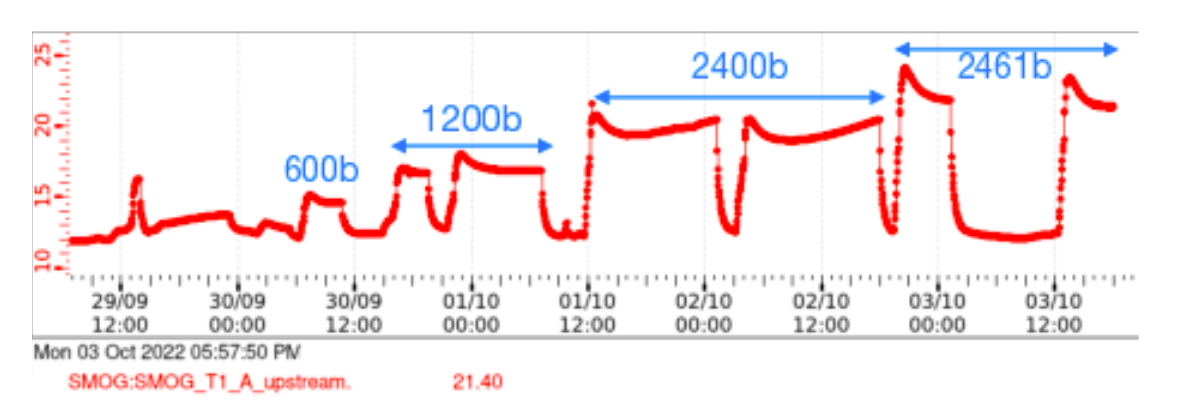} 
    \caption{Location of the five thermocouples monitoring the temperature across the SMOG2 cell. The bottom panel shows the TC1 temperature trending (in Celsius) during subsequent \lhc runs with increasing number of circulating bunches. A clear correlation between the beam intensity and the cell temperature can be seen.}
    \label{fig:temp}
\end{figure}
The temperature of the SMOG2 cell is  monitored  using five K-Type twisted pair thermocouple wires, which are additionally insulated with kapton and whose positions across the SMOG2 cell can be seen in Fig.~\ref{fig:temp}.
The values read by the sensors provide the temperature profile along the cell, affecting the conductance of the injected gas and allowing for the calculation of the integrated areal density of the cell.
These thermocouples are securely attached to custom-made Cu-Be2 ring terminations, which are connected to the cell using sensible spot-welded terminations and bolts that also hold the cell in place.
Of the five terminations, three are positioned on the fixed half of the cell (upstream, center near the gas feeding capillary, and downstream), while the remaining two are located on the floating half (upstream and downstream).
The thermocouple wires are fixed to the gas-feeding capillary pipe until reaching a UHV DN40CF Sub--D15 feedthrough. In the flange, both the in-vacuum and air connections are made using UHV ceramic connectors with aluminum housing. Each wire is terminated with appropriate metal pins on both ends.
To establish the connection from the \velo vessel to the DAQ system, a 25-meter-long halogen-free special cable is used. This cable consists of twisted XLPE insulated pairs with overall screen insulation of polyester tape and aluminum/polyester tape. At the end of the cable, close to the Gas Feed System table, a standalone system based on a compact reconfigurable input/output module\footnote{National Instrument cRIO 5047 plus the NI-9214 card} is used to acquire the data from the thermocouples. The system is controlled remotely via Ethernet.
The thermocouples have been calibrated at three fixed points, melting ice, boiling water, and boiling ethanol, correcting for the atmospheric pressure read in the laboratory.
After calibration, the temperature of the cell can be provided with an uncertainty in the order of 0.1~K at 297.2~K.
The bottom panel of Fig.~\ref{fig:temp} shows a typical temperature monitoring plot by the TC1 probe during subsequent \lhc runs with increasing number of circulating bunches. A clear correlation between the \lhc beam intensity and the cell temperature is found, as expected from the increasing power dissipated in the aluminium cell walls.

\subsection{Alignment}
\label{sec:alignment}

The nominal distance of the cell walls from the beam is 5\mm, which is reduced to 3\mm in case of special runs, like the Van der Meer luminosity scans. A careful and correct alignment of the storage cell is hence mandatory.

The storage cell was adjusted in position and orientation with respect to the \velo RF foil based on geodetic metrology performed by the \cern team responsible for experiments surveying and alignment. Before subsequent alignment procedures, the positions of reference points relative to the main axis of the cell are measured (fiducialisation). The reference marks of the SMOG2 system consist of four 8H7 holes located at the four corners of the cell wing. 
The fiducialisation work was performed in a metrology lab and provided with an accuracy of 100\mum. To identify the position of the SMOG2 mechanical structure, and to determine its azimuthal angle in the \lhcb reference frame, virtual points have been added.
\begin{figure}
    \centering
    \includegraphics[width = \linewidth]{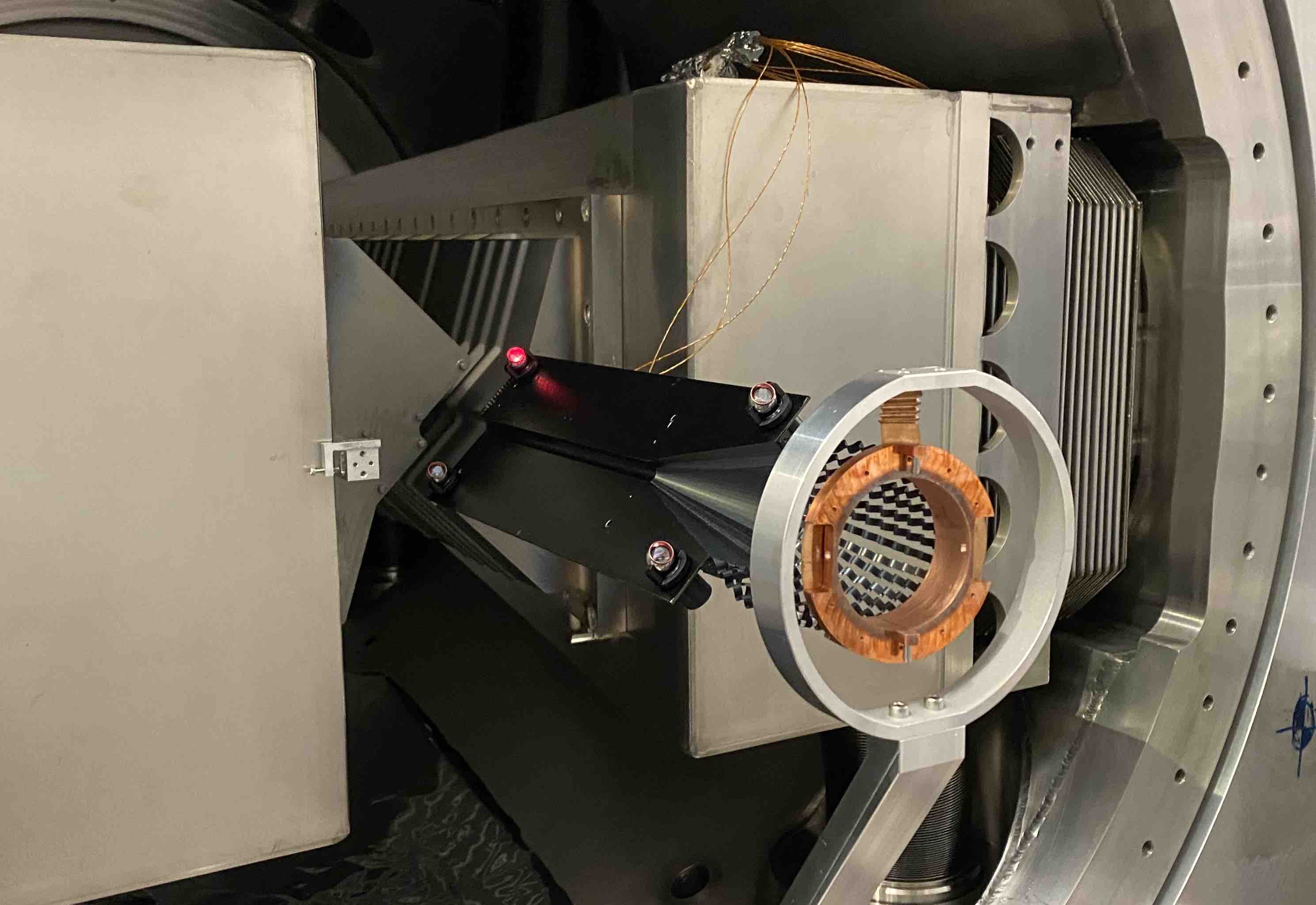}
    \caption{Storage cell fixed on the RF foil and equipped with 4 survey targets in the corners of the wings. One mirror is illuminated by the laser (red spot).}
    \label{fig:refmarks}
\end{figure}
To achieve sub-millimetric precision, a Leica AT402 laser tracker~\cite{lg:2013} was used with corner cubes retro-reflectors in spherical targets, mounted on adaptors with 8g6 pins to fit the SMOG2 fiducial marks, see Fig.~\ref{fig:refmarks}. Multiple alignment sessions were carried out in grey rooms to test and improve the mechanical adjustment system. This was designed to be kinematic to allow stress-free movements, and the lever arms were extended to facilitate angle corrections. After being brought onto the beam, the SMOG2 storage cell was aligned with respect to the previously installed \velo RF foil. The positions and orientations of the \velo RF foil and SMOG2 storage cell were measured from multiple stations, using the \lhcb coordinate system as an intermediate reference frame. The positions of the reference points were determined with a precision of 100\mum at a one-sigma level. The adjustment of the laser tracker stations in the \lhcb \velo alcove was made using \cern 3D adjustment software \cite{barbier:2016}. 
A six degrees of freedom Helmert 3D transformation 
of the fiducialisation data on these points enabled the determination of the positions of the points of interest, and the calculation of the azimuthal angle. The residuals of the transformation were of the order of 250\mum, showing deformation of the wing between the fiducialisation and the final alignment, possibly due to the weight of the survey targets. After several iterations, the adjustments brought the SMOG2 cell 
well below the limits imposed by the maximum beam aperture in that region. The final discrepancies between the measured positions and the nominal positions defined by fiducialisation and the \velo RF foil positions are within 250 $\mu$m in translation
and, 0.51\mrad in azimuthal angle\footnote{The storage cell has a cylindrical symmetry, so the azimuthal angle position is not relevant.}
The position along the beam line is less critical and is defined by the length of the two cantilevers supporting the cell. The final position, measured with respect to the beam-beam interaction point, is currently -536.5~mm for the cell upstream edge and -336.5~mm for the cell downstream edge. The measurement accuracy is 0.2 mm, determined by the uncertainty in the flexible WFS position.

\section{Interference with beam}
%\section{Interference with beam}
\label{sec:beam}

\subsection{Aperture and impedance}
Once the \lhc beams are declared stable for data acquisition, the upgraded \velo detector has in its closed position a minimal nominal distance of 3.5\mm from the beam axis, an aperture that is considered safe in the expected (HL-)LHC conditions of Run~3 and Run~4~\cite{LHCb-PUB-2012-018}. It is worth noting that in nominal condition the aperture is always limited by the downstream part of the RF boxes, Fig.~\ref{fig:cell1b}.
However, several effects were
accounted for, including the transverse offset imposed by the beam crossing configuration, waist shift, beta-beating, and the expected orbit shift during the physics fill. Furthermore, several machine configurations were studied, with baseline optics as well as smaller values of $\beta^*$, both horizontal and vertical crossing 
configurations, and also special runs like $\beta^*$-leveling, ion runs
and van der Meer scans. The studies show that the minimum allowed aperture over the longitudinal range of the SMOG2 storage cell is imposed by the van der Meer scan configuration
and amounts to 3\mm (assuming that the storage cell is centered around the closed orbit at every fill). Given that the storage cell aperture is 5\mm, there is ample space to accommodate these tolerances with a sufficient margin. 

Bunched beams with 40~MHz bunch frequency and high bunch charge represent strong sources of electromagnetic fields. The general rules for guiding these beams safely are: (i) to surround them with conducting surfaces that vary as smoothly as possible in cross-section in order to
keep the RF field close to the beams, and (ii) to avoid excitation of cavity-like structures or other resonating systems. 
Electromagnetic simulations were used to clarify the impact of the WFS system on the \lhc. This consisted of eigenmode calculations,
frequency domain wire simulations, and time-domain wakefield simulations. 
The additional contribution to the low-frequency broadband impedance due to the SMOG2 setup is found to remain small compared to that of the \velo. As a consequence, the
\lhc longitudinal and transverse beam stability is not altered significantly by the addition of the SMOG2 setup. 
Additionally, no evidence has been found that the SMOG2 setup modifies longitudinal and transverse resonant modes in both open and closed positions~\cite{CERN-PBC-Notes-2018-008}.

\subsection{Secondary Electron Yield and coating}

Electron multipacting has been observed in particle accelerators with positively charged beams, leading to the formation of electron clouds that may cause beam instabilities, pressure rise, and heat loads. To avoid any detrimental impact of the storage cell on the \lhc beam dynamics and operation, the cell surface was coated with a low Secondary Electron Yield (SEY) thin film.

Two types of thin films are used at \cern to reduce the SEY in the beam pipes: Ti-Zr-V~\cite{CHIGGIATO2006382, HENRIST200195} and amorphous carbon (aC)~\cite{PhysRevSTAB.14.071001, vollenberg:ipac2021-wepab338}. The low SEY of the Ti-Zr-V (NEG) is achieved after reducing the surface oxides by heating in vacuum at a temperature above 180 $^\circ$C for a few hours (a process called activation). However, after activation, the Ti-Zr-V film pumps hydrogen and other reactive species by gettering effect. As hydrogen is one of the gases to be injected as fixed target, the gettering effect of the Ti-Zr-V film could compromise the stability and reproducibility of the gas density in the cell. Therefore, the aC coating option was chosen,
which exhibits a maximal SEY of one, does not require any activation process, and is inert with respect to the injected gases.

Typical thicknesses of the aC film as used in the \lhc are in the range from 50\nm to 200\nm~\cite{EDMS_1983116}, with a pre-layer of Ti to enhance adhesion (between 100\nm and 200\nm). The aC coating was applied not only to the surfaces facing the beam to avoid electron multipacting, but also to the back of the storage cell, to increase the emissivity of the aluminum surface, enhance thermal exchanges with the surrounding and ease the dissipation of the heat generated by image charge currents in the storage cell.

The Ti pre-layer and the aC film were deposited by Direct Current (DC) magnetron sputtering in a planar geometry, using two 150\mm diameter Magnetron sources (U.S. Inc. Mak), equipped with Ti (grade 2) and graphite targets (Steinemann Carbon AG, R8710, ashes content $<200 \,ppm$). Before launching the coating process, the system was baked in vacuum for 10 hours at 100~$^\circ$C, yielding a base pressure of about 10$^{-5}$~Pa. Argon~40, with a purity of 99.9999\%, was used as discharge gas. A detailed description of the coating system is provided in Ref.~\cite{condmat5010009}.

\begin{figure}
    \centering
    \includegraphics [width = \linewidth] {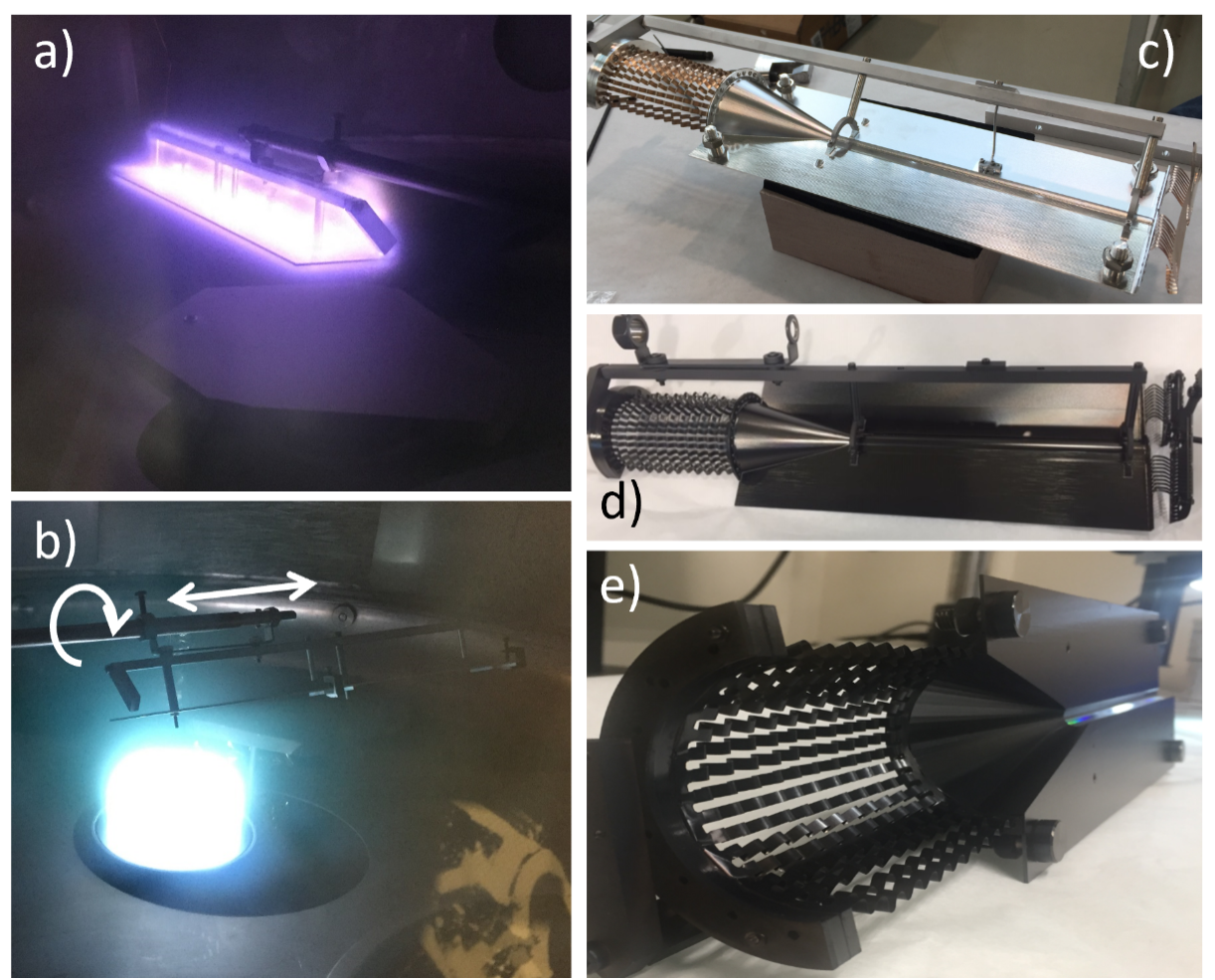}
    \caption{a) DC diode glow discharges during the ion etching step for surface preparation. Argon ions bombarding the surfaces of the cell, removing the oxide layer; b) DC magnetron glow discharge during the deposition process. The Ar ions bombard the targets (Ti or graphite), sputtering the atoms that are deposited on the surfaces of the cell facing the targets; c) Picture of a half-cell assembly before coating; d) and e) a half-cell after coating.}
    \label{fig:pedro1}
\end{figure}

A study was carried out to find the best surface preparation to ensure flawless adhesion on the different materials of the cell. For the aluminum surfaces, it was found that the combination of \cern standard degreasing~\cite{EDMS_1390437} and the Ti pre-layer yielded an adhesion level 0 (very good) by the cross-hatch method (DIN EN ISO 2409). On the flexible Cu-Be2 surfaces of the wakefield suppressors, standard degreasing, followed by in-situ ion etching of the oxide layer (with Ar ions) and the Ti pre-layer proved to yield a flawless adhesion after mechanical cycling ($>100$ bending cycles). To minimize the manipulation of parts after coating, the coating was applied on pre-assemblies of half-cells, Fig.~\ref{fig:pedro1}c, including the aluminum plates, the Cu-Be2 wakefield suppressors, and the supports. The ion etching step was then applied to all parts. Figures \ref{fig:pedro1}a and b show the glow discharge for the ion etching and the aC deposition steps on an Al piece during optimization and calibration.

Given their geometry, the half-cells assemblies were mounted on a shaft combining translation and rotation movements, Fig.~\ref{fig:pedro1}b, to obtain a thickness distribution of the aC film between 50\nm and 200\nm on all the surfaces. The coating parameters for the Ti and carbon layers, including the different longitudinal and angular positions of the cell, the discharge power, voltage and current, and the deposition time, are summarized in Tab.~\ref{tab:pedro1} and Tab.~\ref{tab:pedro2}. The Ar pressure was kept constant at $1.2\times 10^{-1}$~Pa for all the positions. Figure~\ref{fig:pedro1}c shows a half cell before coating, Fig.~\ref{fig:pedro1}d and \ref{fig:pedro1}e after it. Measurements on witness samples yielded maximal SEY of 1.00 and 1.02. 

\begin{table}
    \centering
\begin{tabular}{ |p{1.8cm} p{1.8cm}|p{1.8cm} p{1.8cm} p{1.8cm} p{1.8cm}|  }
\toprule
\multicolumn{2}{|c}{position} &\multicolumn{4}{|c|}{discharge} \\
\multicolumn{1}{|c}{longitudinal}&\multicolumn{1}{c}{angular} &\multicolumn{1}{|c}{power}&\multicolumn{1}{c}{voltage}&\multicolumn{1}{c}{current}&\multicolumn{1}{c|}{coating time}\\
\multicolumn{1}{|c}{[cm]}&\multicolumn{1}{c}{[deg]} &\multicolumn{1}{|c}{[W]}&\multicolumn{1}{c}{[V]}&\multicolumn{1}{c}{[A]}&\multicolumn{1}{c|}{[min]}\\
\midrule
\multicolumn{1}{|c}{}&\multicolumn{1}{c}{0} &\multicolumn{1}{|c}{241}&\multicolumn{1}{c}{268}&\multicolumn{1}{c}{0.9}&\multicolumn{1}{c|}{3}\\
\multicolumn{1}{|c}{}&\multicolumn{1}{c}{30} &\multicolumn{1}{|c}{242}&\multicolumn{1}{c}{269}&\multicolumn{1}{c}{0.9}&\multicolumn{1}{c|}{3}\\
\multicolumn{1}{|c}{0}&\multicolumn{1}{c}{150} &\multicolumn{1}{|c}{242}&\multicolumn{1}{c}{269}&\multicolumn{1}{c}{0.9}&\multicolumn{1}{c|}{8}\\
\multicolumn{1}{|c}{}&\multicolumn{1}{c}{210} &\multicolumn{1}{|c}{243}&\multicolumn{1}{c}{270}&\multicolumn{1}{c}{0.9}&\multicolumn{1}{c|}{8}\\
\multicolumn{1}{|c}{}&\multicolumn{1}{c}{330} &\multicolumn{1}{|c}{243}&\multicolumn{1}{c}{270}&\multicolumn{1}{c}{0.9}&\multicolumn{1}{c|}{3}\\
\midrule
\multicolumn{1}{|c}{}&\multicolumn{1}{c}{0} &\multicolumn{1}{|c}{244}&\multicolumn{1}{c}{271}&\multicolumn{1}{c}{0.9}&\multicolumn{1}{c|}{3}\\
\multicolumn{1}{|c}{130}&\multicolumn{1}{c}{30} &\multicolumn{1}{|c}{244}&\multicolumn{1}{c}{271}&\multicolumn{1}{c}{0.9}&\multicolumn{1}{c|}{3}\\
\multicolumn{1}{|c}{}&\multicolumn{1}{c}{150} &\multicolumn{1}{|c}{244}&\multicolumn{1}{c}{271}&\multicolumn{1}{c}{0.9}&\multicolumn{1}{c|}{8}\\
\bottomrule
        \end{tabular}
  \caption{Parameters used for the coating process with Titanium.}
    \label{tab:pedro1}
\end{table}
\begin{table}
    \centering
\begin{tabular}{ |p{1.8cm} p{1.8cm}|p{1.8cm} p{1.8cm} p{1.8cm} p{1.8cm}|  }
\toprule
\multicolumn{2}{|c}{position} &\multicolumn{4}{|c|}{discharge} \\
\multicolumn{1}{|c}{longitudinal}&\multicolumn{1}{c}{angular} &\multicolumn{1}{|c}{power}&\multicolumn{1}{c}{voltage}&\multicolumn{1}{c}{current}&\multicolumn{1}{c|}{coating time}\\
\multicolumn{1}{|c}{[cm]}&\multicolumn{1}{c}{[deg]} &\multicolumn{1}{|c}{[W]}&\multicolumn{1}{c}{[V]}&\multicolumn{1}{c}{[A]}&\multicolumn{1}{c|}{[min]}\\
\midrule
\multicolumn{1}{|c}{}&\multicolumn{1}{c}{0} &\multicolumn{1}{|c}{431}&\multicolumn{1}{c}{718}&\multicolumn{1}{c}{0.6}&\multicolumn{1}{c|}{20}\\
\multicolumn{1}{|c}{}&\multicolumn{1}{c}{30} &\multicolumn{1}{|c}{430}&\multicolumn{1}{c}{717}&\multicolumn{1}{c}{0.6}&\multicolumn{1}{c|}{5}\\
\multicolumn{1}{|c}{0}&\multicolumn{1}{c}{150} &\multicolumn{1}{|c}{462}&\multicolumn{1}{c}{770}&\multicolumn{1}{c}{0.6}&\multicolumn{1}{c|}{35}\\
\multicolumn{1}{|c}{}&\multicolumn{1}{c}{210} &\multicolumn{1}{|c}{455}&\multicolumn{1}{c}{759}&\multicolumn{1}{c}{0.6}&\multicolumn{1}{c|}{35}\\
\multicolumn{1}{|c}{}&\multicolumn{1}{c}{330} &\multicolumn{1}{|c}{430}&\multicolumn{1}{c}{716}&\multicolumn{1}{c}{0.6}&\multicolumn{1}{c|}{5}\\
\midrule
\multicolumn{1}{|c}{}&\multicolumn{1}{c}{0} &\multicolumn{1}{|c}{428}&\multicolumn{1}{c}{713}&\multicolumn{1}{c}{0.6}&\multicolumn{1}{c|}{20}\\
\multicolumn{1}{|c}{130}&\multicolumn{1}{c}{30} &\multicolumn{1}{|c}{430}&\multicolumn{1}{c}{716}&\multicolumn{1}{c}{0.6}&\multicolumn{1}{c|}{5}\\
\multicolumn{1}{|c}{}&\multicolumn{1}{c}{150} &\multicolumn{1}{|c}{464}&\multicolumn{1}{c}{773}&\multicolumn{1}{c}{0.6}&\multicolumn{1}{c|}{35}\\
\bottomrule
        \end{tabular}
        \caption{Parameters used for the coating process with amorphous carbon.}
    \label{tab:pedro2}
\end{table}

\subsection{Simulation for coating saturation}
\label{sec:Molflow}
After a number of wall bounces, the gas molecules injected into the storage cell exit from one of the two ends and, in case of non-noble gases, can interact with the NEG coating of the beam pipe and \velo RF boxes, impacting its performance and possibly inducing peel off. It is thus important to understand the magnitude of the gas flow and its propagation outside of the storage cell, also in order to set realistic flux and injection time limits to the operation of the gas feed system. As mentioned in Sec.~\ref{sec:sc_principle}, \textit{Molflow+}, a molecular flow Monte Carlo simulator, can be used to determine the pressure profile and the gas propagation in an arbitrarily complex geometry. 
\textit{Molflow+} simulates the collisions (hits) of the molecules with each surface in the implemented geometry (facets), characterized by a unique temperature, opacity, and sticking coefficient, which is defined as the probability that an impinging particle gets captured by the NEG coating. It is a steady-state simulator, which means that during the simulation there is a continuous gas flow rate Q with constant parameters on the facets. For an ideal gas of pressure \textit{p}, volume V and temperature T, the flux rate of  particles entering the system $dN_{real}/dt$ is given by:
\begin{equation}
        \label{eq:mol}
        \begin{split}
            & Q = \frac{d\left(pV\right)}{dt},\\
            & \frac{dN_{real}}{dt} = \frac{d\left(pV\right)}{dt}\frac{1}{k_{B}T} = \frac{Q}{k_{B}T},
        \end{split}
    \end{equation}
being $k_{B}$ the Boltzmann constant and where the ideal gas law has been used. \textit{Molflow+} applies the test-particle Monte Carlo method: a limited number of virtual test particles $N_{virtual}$ is generated to represent a larger rate of physical molecules through a determined scale factor $K_{r/v}$:
\begin{equation}
 K_{r/v} = \frac{dN_{real}}{dt}/N_{virtual}.
\end{equation}
Quantities depending on the rate, such as absorption rate, pressure or particle density are directly simulated; the absolute ones are obtained by multiplying the rates by the physical time of interest. A dedicated workflow was implemented for time-dependent simulation allowing a dynamic evolution of the facet parameters. It relies on iteratively calculating and updating the facet parameters after a short step of time, through the \textit{Molflow}CLI, rather than running a single time dependent simulation over a long time frame but with static facet parameters. The time-dependent simulation mode already implemented in the \textit{Molflow+} graphical user interface does not allow to update the facets parameters during the simulation based on the simulation results themselves. The command-line interface, allowing an automated control of the workflow via configuration files, can be exploited instead to implement a dynamic parameter evolution. 

Non-noble gases can be categorized based on their interaction with the surface coated with NEG: 
\begin{itemize}
    \item getterable gases, like $N_2$ and $O_2$, tend to stick to the surface, reducing the free adsorption sites and thus the effective NEG sticking coefficient over time. While the reduced pumping speed does not impact the \lhcb experiment operation, the NEG saturation can produce a detrimental SEY increase;
    \item Hydrogen-like gases dissociate on the NEG surface and diffuse into the bulk. While the sticking coefficient depends weakly on the surface concentration, increased bulk concentrations can induce embrittlement and NEG peel off.
\end{itemize} 
For both types of gas, precise evaluation of the expected impacts on the RF box NEG is a prerequisite to proceed with the injection. To achieve this, laboratory measurements were carried out on NEG samples and, at the same time, a dynamic flow simulation was implemented in \textit{Molflow+} to predict the time evolution of the saturation. Considering that the impact on the NEG is higher the closer to the injection point, the simulated geometry was limited to the storage cell and the \velo RF box, characterized by a corrugated surface that makes the understanding of the flow inside it a non-trivial problem.

H$_{2}$ and N$_{2}$ were considered as proxies for hydrogen-like and getterable gases, considering $100$ and $10$~hours of injection at $1.5\times10^{-4}$ and $4.05\times10^{-5}$~mbar l/s,  respectively. A fit to the available sticking coefficient experimental data (see Ref.~\cite{CHIGGIATO2006382} for N$_{2}$ and Sec.~\ref{sec:h2} for H$_{2}$) was performed to parametrise the sticking coefficient evolution as a function of the gas surface concentration.

\begin{figure}
	\centering
\includegraphics[width=\linewidth, clip]{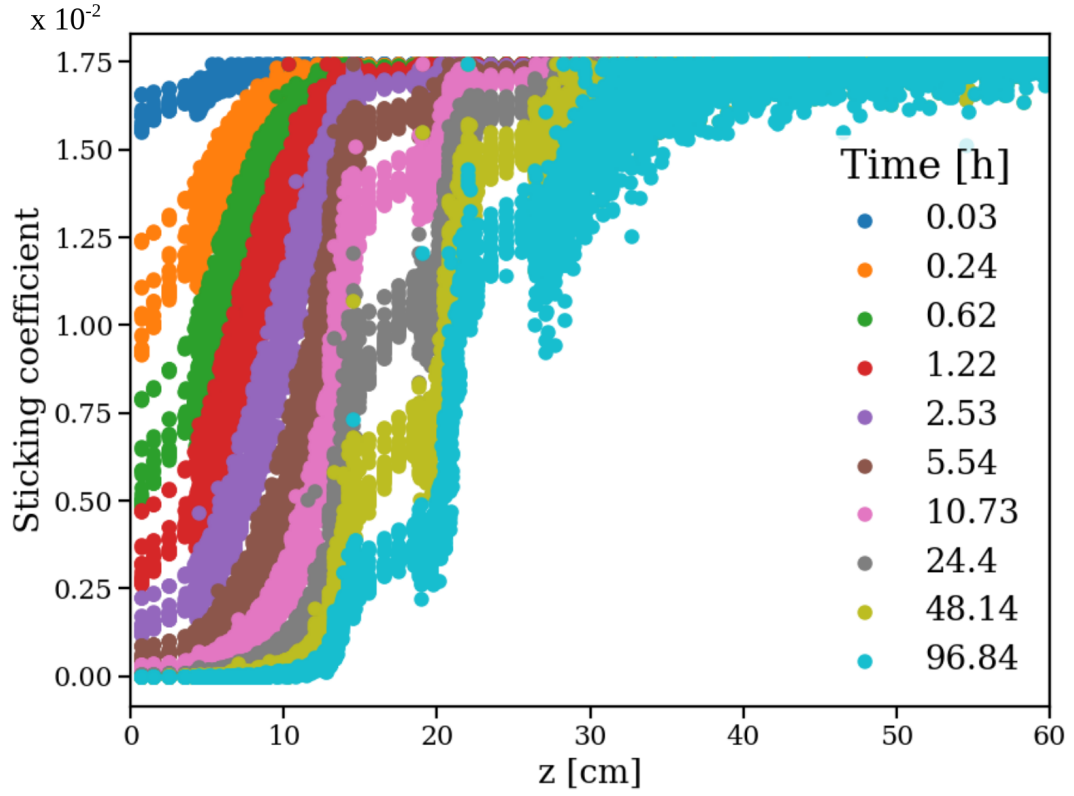}
\includegraphics[width=\linewidth, clip]{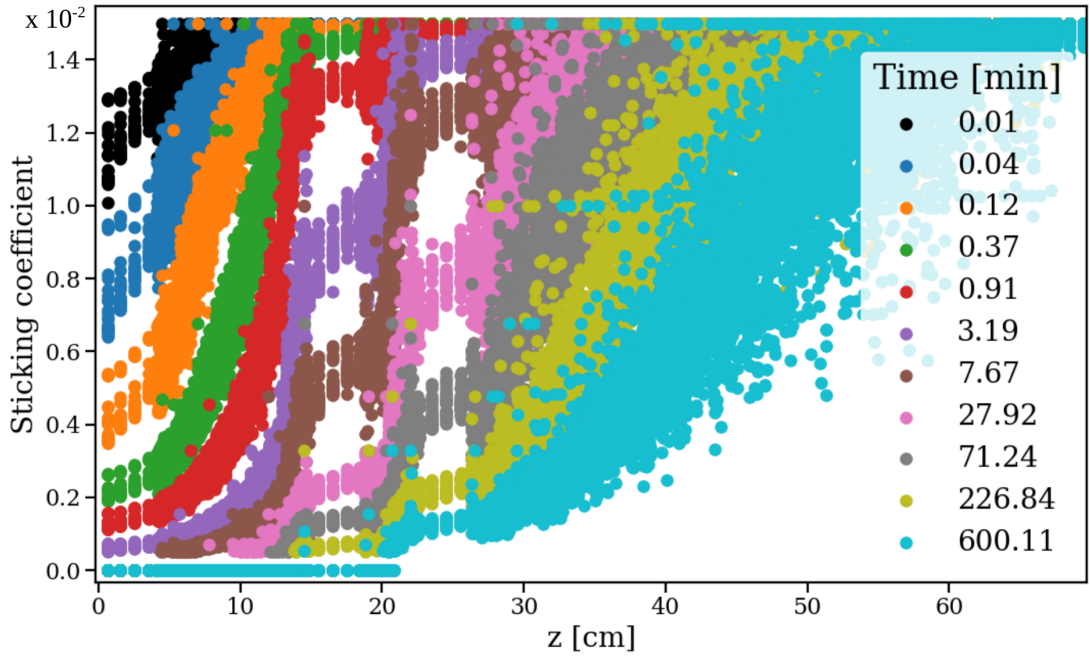}	
\caption{RF foil sticking coefficient as a function of the longitudinal position $z$ for different periods of injected H$_{2}$ (top) or N$_2$ (bottom). Each dot represents the sticking coefficient of a facet, with $z$ indicating the longitudinal coordinate of its center and being $z=0$ the upstream boundary of the RF foil.}
	\label{fig:molH2}
\end{figure}

Figure ~\ref{fig:molH2} presents the results of the simulated sticking coefficient evolution with the \textit{z} coordinate and the injection time. On top, for H$_{2}$, the onset of saturation begins after around $20$ hours of injection and, after $100$~hours, the saturation only reaches the central region of the first 100\mm of RF foil, corresponding to around $2\%$ of the total area. On the bottom, for N$_{2}$, the saturation onsets after the first minutes from the injection and progresses much faster, as expected, reaching up to $200$~mm, corresponding to more than $15\%$ of the total area, after $10$~hours. 

The results of these simulations, together with direct measurements on NEG samples in laboratory, assured that the level of saturation prospected in the RF box during Run 3 operation causes no safety issue to the \lhc operations.

%Giuseppe B, Josef, David
\subsection{\texorpdfstring{H$_{2}$}{H2} injection}
\label{sec:h2}
%1 page

In order to understand and evaluate the potentially detrimental effects of H$_{2}$ injection on the NEG coating with regard to saturation and embrittlement, a laboratory scale H$_{2}$ saturation study was performed on 2\m-long and 3.5\cm-diameter NEG coated stainless steel beam pipes. Two injection pressure conditions were considered, (i) 1~mbar H$_{2}$ injection pressure, to test the H$_{2}$ embrittlement limit of the NEG coating, and (ii) 5$\times 10^{-7}$~mbar H$_{2}$ injection pressure to simulate its saturation under conditions similar to the SMOG2 data-taking ones~\cite{DMP_thesis}.

\begin{figure}
	\centering
\includegraphics[width=0.9\linewidth, clip]{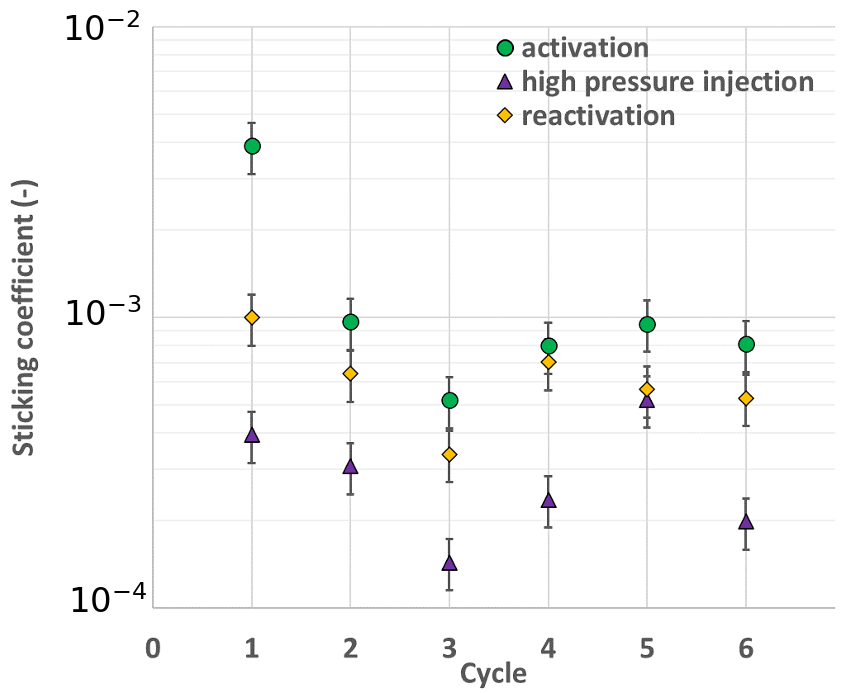}
\includegraphics[width=0.9\linewidth, clip]{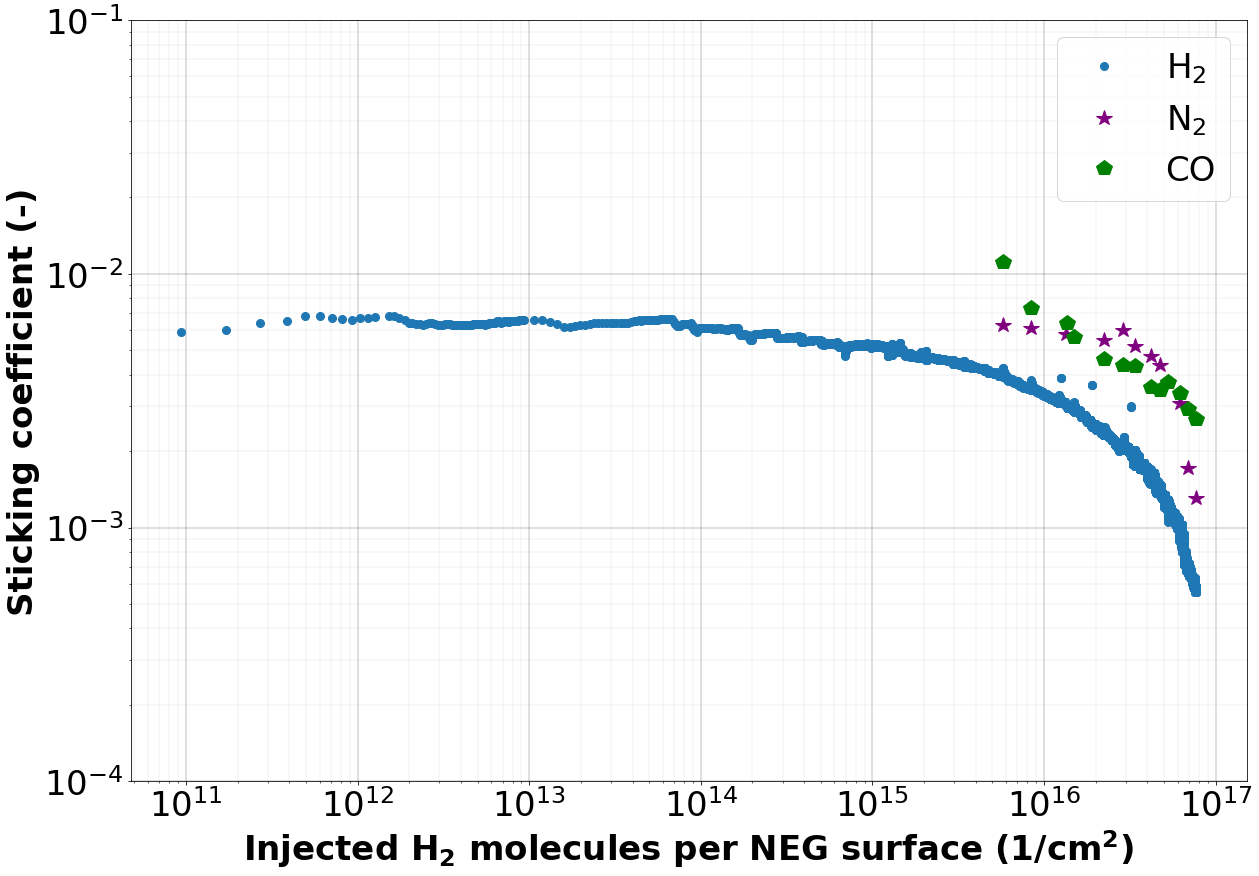}
	\caption{Experimentally obtained H$_{2}$ sticking coefficient evolution of NEG coating throughout saturation with H$_{2}$ gas under quasi-instantaneous 1 mbar (top), and continuous 5$\times 10^{-7}$ mbar (bottom) injection pressure conditions~\cite{DMP_thesis}.}
	\label{fig:H2_sat_study}
\end{figure}

The saturation experiment performed at 1~mbar injection pressure consisted of 6 cycles and in each cycle the NEG coating was subjected to two consecutive quasi-instantaneous injections of H$_{2}$ gas at a nominal pressure of 1~mbar. As the NEG coating was vented to air between the cycles, an activation was performed at the beginning of each cycle, as well as at the end of each which was to study how replenishable the sticking coefficient -- measured with the transmission method~\cite{10.1063/1.2436092} -- is after saturation with H$_{2}$. The sticking coefficient evolution of the NEG coating throughout the 6 cycles is shown in the top plot of Fig.~\ref{fig:H2_sat_study} as measured after the initial activation, after the first H$_{2}$ injection at 1~mbar, and after the reactivation. The saturation experiment at an injection pressure of 5$\times 10^{-7}$~mbar was carried out with continuous injection, during which the measured evolution of the H$_{2}$ sticking coefficient on the NEG coating was recorded, as shown in the bottom plot of Fig.~\ref{fig:H2_sat_study}. In addition, N2 and CO were used for comparison. It is important to note that a trend similar to that of H2 is observed.

Saturation of the NEG coating was reached under both injection pressure conditions, namely at the minimum calculated concentrations of 0.407 H/TiZrV mol/mol at the 1~mbar injection condition and 0.026 H/TiZrV mol/mol at the 5$\times 10^{-7}$~mbar injection pressure condition. While the decrease of the H$_{2}$ sticking coefficient of the NEG coating was observed following its saturation, signs of embrittlement of the NEG coating were not found by endoscopic analysis of the test pipes, indicating safe operability of the NEG coating up to the tested injection condition pressures and quantities.

\subsection{Machine Induced Background}

The evaluation of the amount and characteristics of backgrounds induced by the beam circulation is both relevant to the design of the structure and the understanding of possible degradation of the running conditions.

The approach used in this study follows the methodology described in Refs.~\cite{Appleby:1376692, LHCb-INT-2011-015}, where numerical analyses of the Machine Induced Background (MIB) at \lhc have been implemented. 
Specifically, the analysis took into account the interactions between the proton beam and residual gas or nearby materials along the beamline, such as long straight sections and tertiary collimators, in the presence of the storage cell.
Simulations demonstrate that the inclusion of the storage cell system's material budget in front of the \lhcb detector has no impact on the number of \velo clusters per event in \pp collisions. The MIB alone leads to a maximum absolute variation of +16\%. However, when appropriately scaled and embedded into the \pp collisions, the storage cell influence becomes completely negligible.

With injected gas, an additional mechanism of beam loss arises due to beam-gas collisions. 
The impact on the beam lifetime can be described in terms of the total beam-gas
cross-section $\sigma_{loss}$.
Considering 
\begin{equation}
    \sigma_{loss}\simeq \sigma_{pN}\simeq A^{2/3}\sigma_{pp},
\end{equation} with $\sigma_{pp}\simeq$ 50 mb and $A$ the mass number of the considered gas, the expected beam lifetime amounts to 2060, 97, and 22 days for \pH, \pAr and PbAr,  respectively, largely exceeding the typical duration of a \lhc fill of 10-12 hours.

\subsection{Heating by the beam}
To assess possible heating effects due to pick-up from the beam RF, heating tests and simulations have been conducted using various power levels. These tests were performed on an isolated, coated, closed cell under vacuum conditions. The results indicate that at a power of 15 W, the temperature of the cell reaches 75 $^\circ$C ~\cite{LHCb-TDR-020}.
Furthermore, tests were conducted to heat the cell up to 130 $^\circ$C, and no observable changes in the shape of the cell were detected.
 
\section{The Gas Feed System}
\label{sec:gfs}
\begin{figure*}
\centering
\includegraphics[width=\linewidth]{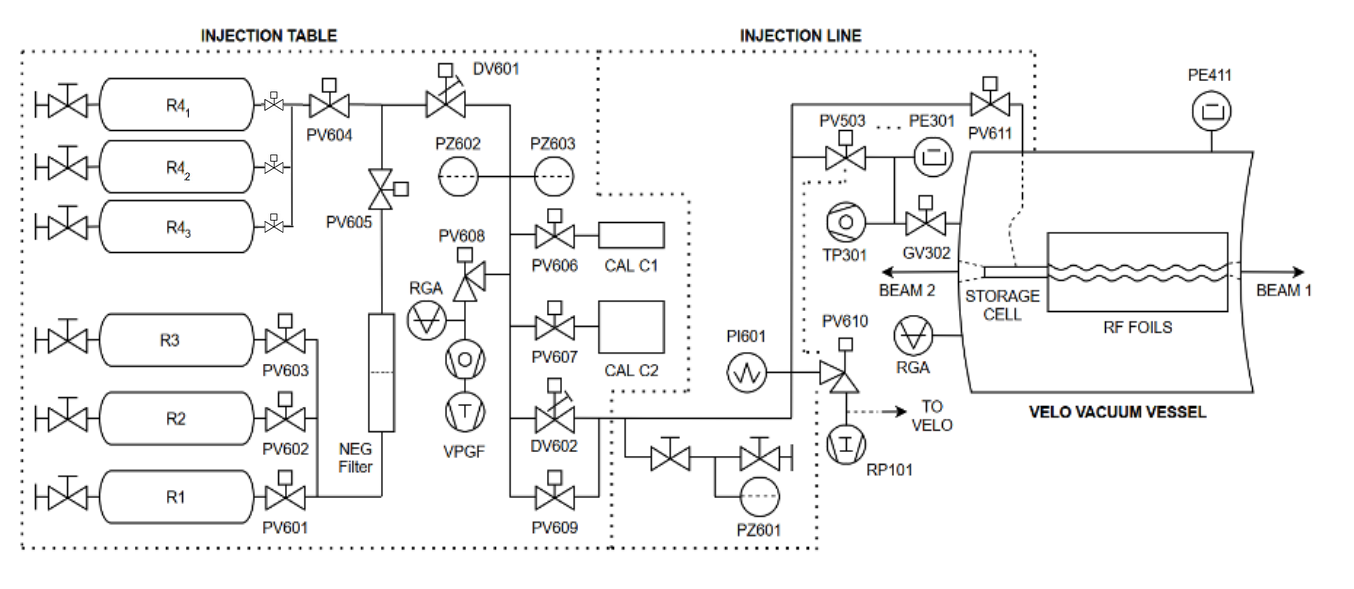}
\caption{General design of the GFS, with the main components indicated for the injection table (left) and the injection line (right). The valves PV601 and PV604 are Swagelok 6BG all-metal valves; PV602, PV603, PV606 and PV607 are Swagelok 6BK valves with polychlorotrifluoroethylene stem. Both DV601 and DV602 dosing valves are Pfeiffer EVR 116. The PZ602 and PZ603 gauges are Pfeiffer CMR 373 and CMR 371 ceramic membrane gauges, PZ601 is Pfeiffer CMR 375 and PI601 a Pfeiffer
Pirani gauge TPR 018. The nominal pumping group VPGF is composed of a turbo-molecular pump Pfeiffer HiPace 80, a primary vacuum pump Edwards RV12 Rotary and a MKS HPQ3 RGA. The turbo-molecular pump TP301 is a Pfeiffer HiPace 700; the primary vacuum pump RP101 a Pfeiffer ACP28.}
\label{fig:GFS1}
\end{figure*}
The Gas Feed System (GFS), allowing precise measurements of the gas flow rates and remote flow adjustments, is composed of an injection table and an injection line, as shown in Fig.~\ref{fig:GFS1}.
While the GFS employed for the previous SMOG system was located in the \velo alcove, the new one is relocated in the \lhcb cavern at a distance of approximately 22\,m from the injection point. Such a position was defined based on the following factors:
\begin{itemize}
\item Available floor space for the GFS injection table $\sim$ 1\,m$^2$. This assumes permanent installation of the device without presenting an obstacle for any coactivation or transport;
\item Tolerable radiation for electronic components installation and operation;
\item Low magnetic field by the \lhcb spectrometer dipole for continuous operation of primary and turbo-molecular pumps;
\item Availability of services, cabling, and routing, such as the supply of compressed air for the valve operation, the routing of the injection line from the \velo alcove to the GFS table, the cabling for the GFS control crates.
\end{itemize}

\subsection{The GFS injection table and injection line}
The injection table hosts the gas reservoirs and all the equipment that is needed for the gas injection preparation and monitoring. While sensitive components such as capacitive gauges, valves and RGA should not exceed 100$^{\circ}$C, all other elements withstood a 24h bake-out at 150$^{\circ}$C to remove water, except for the NEG filter which was activated at 400$^{\circ}$C. Permanent bake-out heaters are also available, allowing bake-out cycles to be performed during technical stops or shutdowns.

All the gas reservoirs store  up to a 1.5~bar pressure in a 1~liter volume and are equipped with an interface for the gas refilling. The R1, R2, and R3 reservoirs are designated for noble gases, and are connected with noble gas purification bypass (SAES purifier PS10) via the process valves PV601, PV602 and PV603, as indicated in Fig.~\ref{fig:GFS1}.  The injected gas is purified with a commercial NEG filter, which reduces the concentration of contaminants from ppm to ppb level~\cite{Bregliozzi_NEG}. The reservoirs R4 are dedicated to store getterable gases and are directly connected to a high-pressure feeding arm via PV604.

All the injected gas is dosed to a low-pressure measurement arm using the variable leak valve DV601. To store a precise quantity of injected gas, two calibrated volumes C1 (0.0565 l) and C2 (0.1565 l) are installed and connected by the process valves PV606 and PV607. The PZ602 and PZ603 gauges allow performing precise measurements of the pressure independently of the gas type, in a range from $1\cdot 10^{-3}$ to 11\,mbar and from 0.1 to 1100\,mbar, respectively. On these measuring volumes, a fixed pumping group composed of a turbo-molecular pump and a primary vacuum pump allows for recovery of the GFS after the injection and for the injection table conditioning during the gas preparation processes. It also contains an interface for helium-leak detection. 

A MKS HPQ3 high-pressure Rest Gas Analyzer (RGA) is also installed on the manifold between PV608 and the turbo-molecular pump, allowing to measure on the GFS the purity of the injected gas through RGA analyses. It can be only operated during technical stop periods, as the control electronics are dismounted during \lhc operations. A second RGA on the \velo vessel is used instead. 

The injection line is routed from the GFS injection table to the \velo vacuum system interface through 11 long tubular manifolds with 10~mm internal diameter fitted with Swagelok VCR connectors joined by flexible elements (Swagelok VCR bellows), for a total length of $\simeq$ 22\m. The final injection is performed through the dosing valve DV602, while the process valve PV609 is used as a bypass if more important conductance is needed, such as for pump-down or the bakeout of the injection. The PZ601 and PI601 gauges are installed 0.4\m and 21.9\m downstream of the injection table and are used to monitor the injection line pressure. In the \velo alcove, the line splits into three branches. The first one connects with the process valve PV503, located on the manifold between the gate valve GV302 and the inlet of the turbo-molecular pump TP301. The second one, allowing injections in the SMOG2 storage cell, ends at the process valve PV611 located on the top of the \velo vessel. The third branch finally connects the injection line with the \velo primary pump RP101 via the process valve PV610.

\subsection{The injection process}
A GFS control process is integrated into the \velo vacuum control. The process consists of i) change of the \velo vacuum system regime; ii) preparation of the GFS table; iii) gas injection control and stabilization.

Firstly, the \velo vacuum system regime needs to be changed from nominal to SMOG. The former is used during \lhcb standard operations, when there is no need of gas injection. The two \velo ion pumps IP431 and IP441 (Agilent VacIon Plus 500) are running and the turbo-molecular pump TP301 is isolated from the beam vacuum system via the closed gate valve GV302. The nominal regime is considered as a safe-state as a potential malfunction of turbo-molecular or primary pump does not affect the state of the \lhc beam vacuum system. During the SMOG regime, instead, the ion pumps are switched off, the turbo-molecular pump is set at nominal state and the gate valve GV302 is open to evacuate the \velo vessel. The transfer between these two regimes is controlled by the operator when there are no beams circulating in \lhc.

Secondly, the gas injection type has to be chosen as:
\begin{itemize}
    \item in the \velo vessel, like for the previous SMOG system. In this case, the gas is injected from the back of the vessel, at approximately $z \simeq \rm{-}750$\mm, and a uniform pressure increase around the \lhcb nominal interaction point with a pressure around $10^{-7}$~mbar~\cite{LHCb-PAPER-2018-031} is obtained;
    \item in the SMOG2 cell, resulting in a more localized pressure bump and thus higher aerial density. 
\end{itemize}

The gas preparation main purpose is to prepare the selected gas for injection, purge the GFS or recover it from an undefined state. The injection line and injection table are pumped down. Then, the selected gas is injected and pumped three times into the table while the injection line stays on static vacuum. The injection process can then start. The high-pressure arm is kept pressurized with the selected gas and an expiration timer for the gas is reset to last 14 days. This is to avoid a potential purity issue due to long-term storage of active gas within the high-pressure part of the injection table. 

As the long injection line requires a long time for the stabilization of the gas injection, a line rapid-fill procedure was developed and implemented for injections in the SMOG2 cell. The injection line bypass PV609 is closed, and the GFS table is prefilled using DV601 with the C1 volume opened with a defined pressure of active gas typically of $9\cdot 10^{-1}$ mbar. This gas is then injected on the injection line via the PV609 bypass valve and held at this state for 5 minutes or until the pressure stabilizes. In the meantime, the injection line is separated by closing PV609 and the injection table is quickly pumped down. Afterwards, the table is re-filled with the selected gas using DV601 up to 10~mbar. The injection in the storage cell can then start by opening the injection valve PV611, followed by the opening of the dosing valve DV602 at a fixed setpoint. Once the injection is completed, the DV602 is closed, followed by the injection valve PV611.

A standard injection in the \velo vessel consists instead of two preparatory and an injection control step. During the first step, the injection line actively pumping is swapped from fixed pumping group on the injection table to turbo-molecular pump TP301. The low-pressure measurement arm is prepared by prefilling via DV601 10~mbar of active gas into the calibrated volumes C1 and C2. A one hour time-out is then activated to allow the operator to start the injection, which happens through the PV503 and GV302 valves. This is a safety constrain to avoid any potential issue with both \velo and GFS. Once expired, the system no longer allows starting the injection and requires the recovery step. The injection is then started; the dosing valve DV602 is opened and in approximately 15 min a stable injection pressure in the vessel is reached.

\begin{figure*}
\centering
\includegraphics[width=\linewidth]{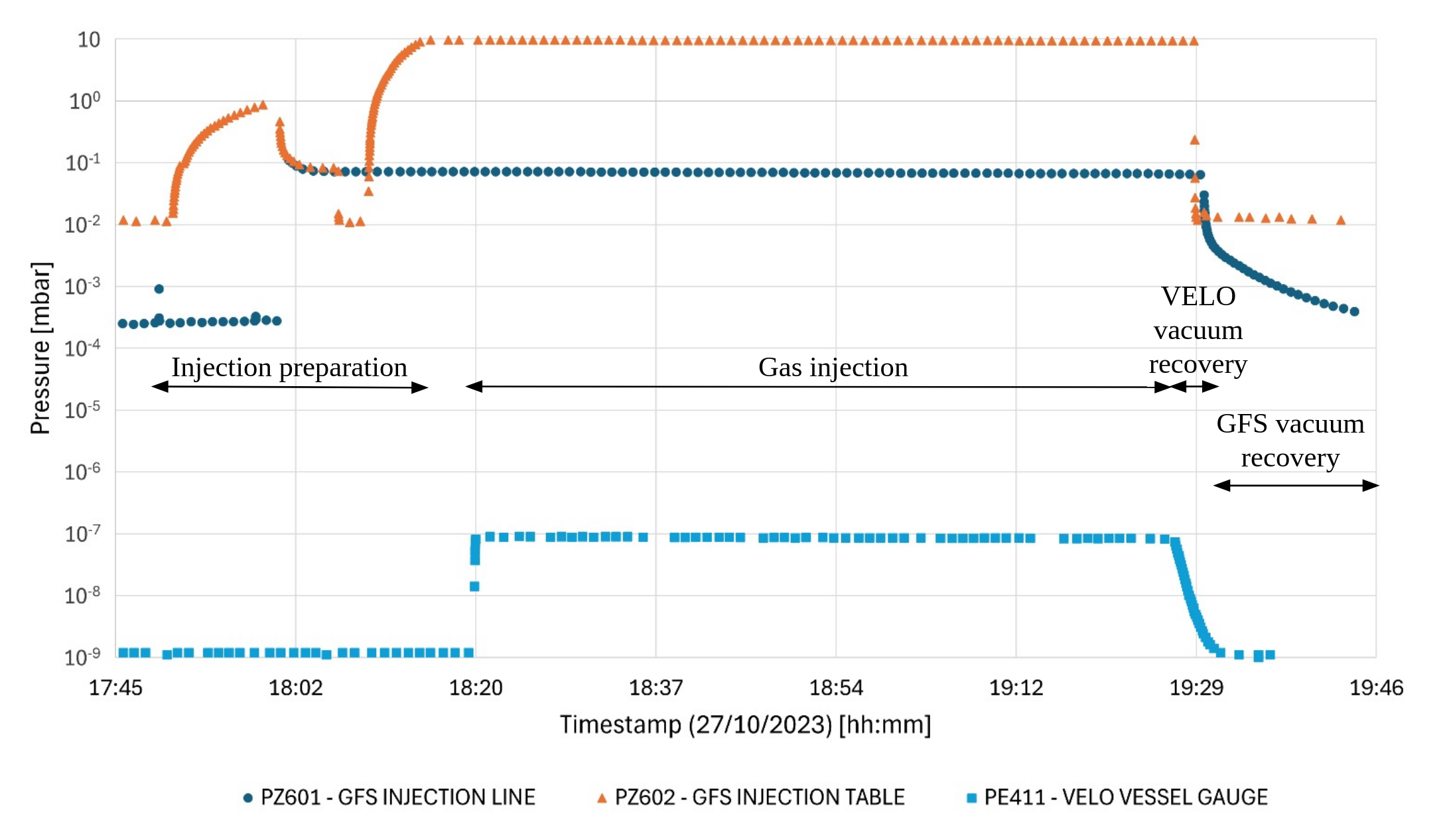}
\caption{Evolution of the pressure readings by the PZ601 (dark blue circles, on the GFS injection line), the PZ602 (orange triangles, on the GFS injection table) and the PE411 (light blue squares, in the \velo vessel) gauges during a standard SMOG2 injection.}
\label{fig:GFS5}
\end{figure*}
\begin{figure*}
\centering
\includegraphics[width=\linewidth]{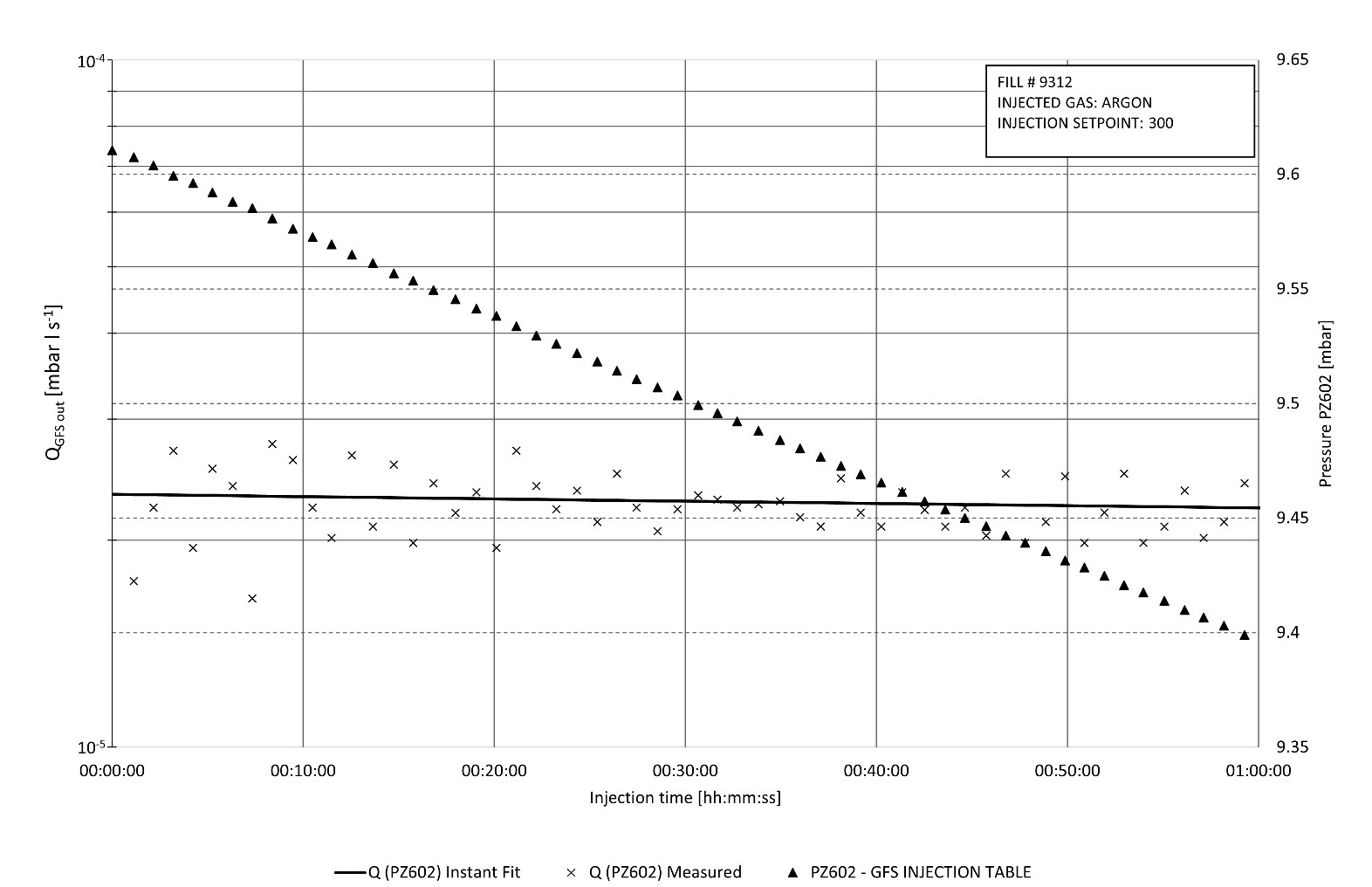}
\caption{Readings from the PZ602 gauge on the GFS table during an injection in the SMOG2 cell and estimation of the instantaneous flux according to the procedure described in the text.}
\label{fig:GFS7}
\end{figure*}

The evolution of various gauge readings during a typical SMOG2 injection is illustrated in Fig.~\ref{fig:GFS5}, with the preparation, injection and recovery steps highlighted. By fitting the readout values of the PZ602 gauge with an exponential function and knowing the C1 and C2 volumes, the flow rate from the GFS table ``Q (PZ602)-Measured" is established, as shown in Fig.~\ref{fig:GFS7}. The instant flow rate fit ``Q (PZ602)-Instant fit" can be used, known the opening of the DV602 valve, to estimate the flux from the reading by the PZ602 gauge during the SMOG2 injection. The stability of the injection in time shows a reduction of the initial flow rate by approximately 4\% per hour assuming the injection pressure on the table is 10~mbar. However, the variation on a few seconds scale, which is relevant for the beam-gas luminosity determination, is below the per-cent level. The precision of the PZ602 readout is mainly affected by the long cables ($\simeq$ 80 m) that connect the gauge head to the controller that is placed in the \lhcb cavern and partially also by the electronics acquisition rate.

Once the ongoing injection, either in the \velo vessel or in the SMOG2 cell, is terminated, the system enters a recovery state to re-estabilsh a \velo beam vacuum pressure lower than $10^{-8}$~mbar. Once this is reached, typically in about 2 minutes, the system proceeds with purging the low-pressure measurement arm of the injection table and setting the GFS system back to its stand-by state. The recovery is finished after 30 minutes, when a new injection, the change of the active gas or the transfer of the vacuum regime back to the nominal one can take place.

\section{Simulations and data-taking preparation}
\subsection{Beam-gas simulated samples}
\label{sec:simulations}
To assess the performance for beam-gas and beam-beam collisions concurrent reconstruction, a set of simulated samples was produced~\footnote{In these simulation, the initial nominal position of the SMOG2 cell, $z \in [-500, -300]$\mm, was considered, shifted with respect to the installation one by 4.1\cm towards the nominal \lhcb interaction point. The conclusions discussed in the following remain valid.} with:
\begin{itemize}
	\item standalone \pp collisions in nominal Run 3 conditions, \ie an average per-bunch crossing number of collisions in \lhcb of $\nu \simeq 7.6$;
	\item standalone \pHe collisions, mimicking the data-taking strategy adopted in 2016 with periods dedicated to collect beam-gas data only;
	\item overlapped \pp and \pHe or \pAr collisions, to exemplify concurrent data acquisition with lighter or heavier gas species.
\end{itemize}
 In all simulated samples, the gas is distributed uniformly in \textit{x} and \textit{y}, while follows a triangular shape in \textit{z}, as discussed in Sec.~\ref{sec:sc}. The contributions due to the small gas flow outside the cell  are expected to only apply negligible corrections to the results discussed in the following.

\subsection{Real-time beam-gas data reconstruction and selection}

In order to cope with the \lhc Run 3 \pp luminosity, $\lum = 2 \cdot 10^{33}\cm^{-2}s^{-1}$, \lhcb completely revisited its data processing strategy. The first and hardware-based data selection level, used until 2018, was discarded and the full detector readout, calibration and alignment, and event reconstruction and selection are now occurring in real-time within a software-only framework~\cite{LHCb-TDR-016}. This is composed of two levels, with a large disk buffer in between to allow detector alignment and calibration constants update:
\begin{figure}
	\centering
	\includegraphics[width = \linewidth]{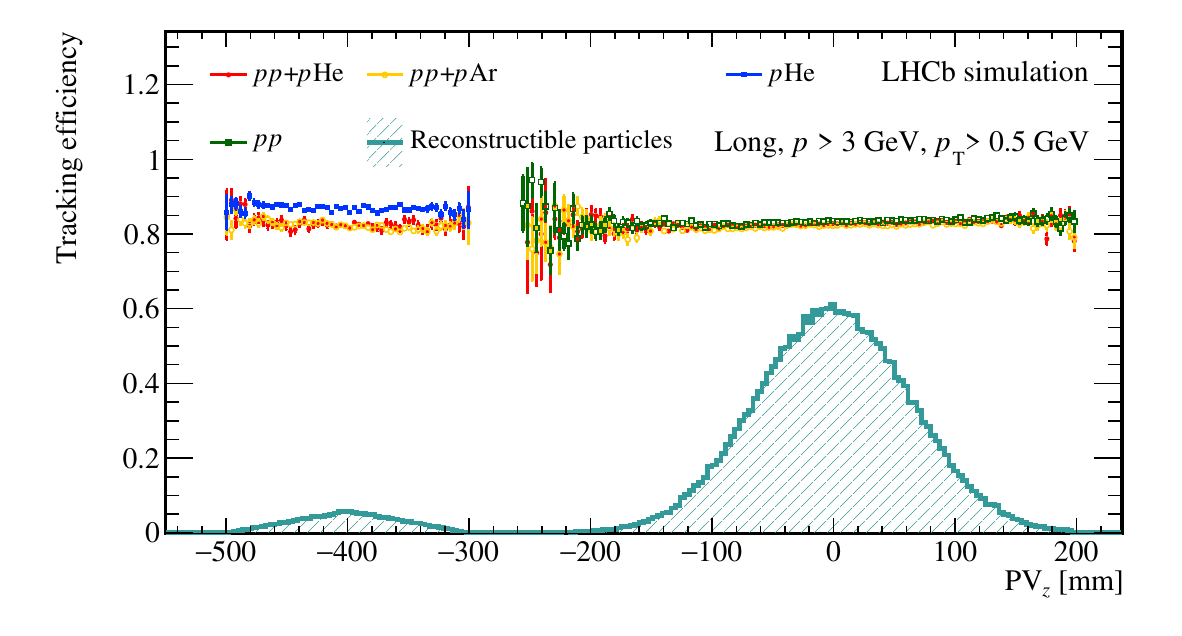}
	\includegraphics[width = \linewidth]{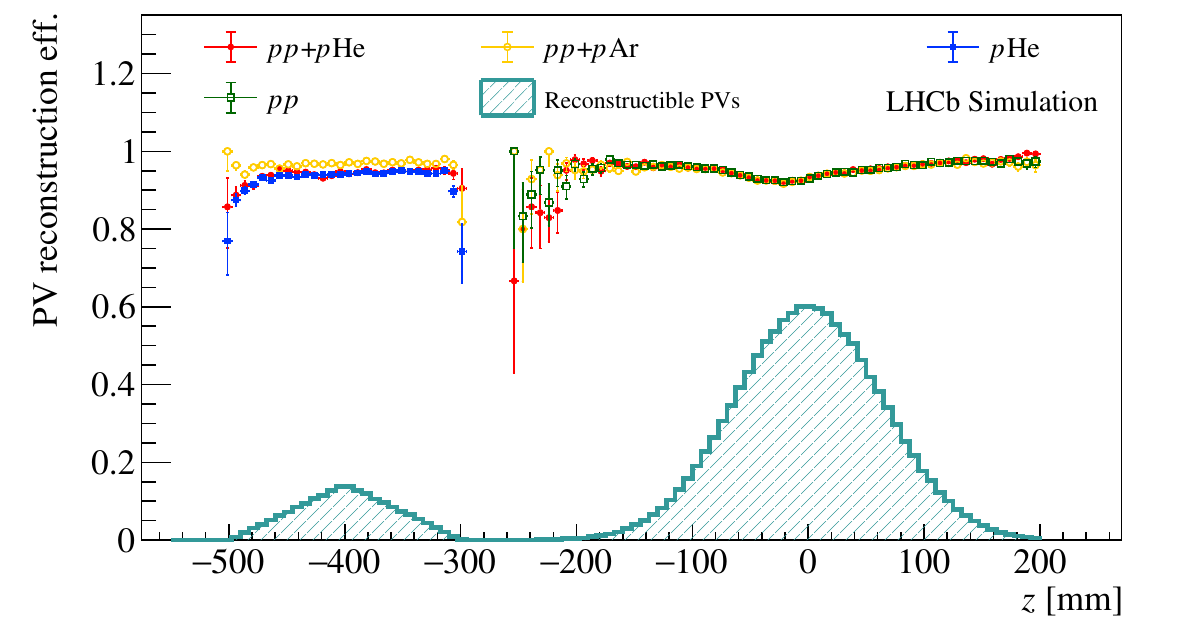}
	\caption{Particle (top) and collision vertex reconstruction (bottom) efficiency as a function of the beam axis longitudinal coordinate, as measured on simulated samples with \pp stand-alone (green), \pHe stand-alone (blue), \pppHe (red) and \pppAr (yellow) collisions. The \textit{z} coordinate distributions for the vertices associated to reconstructible particles (top) or of the vertices themselves (bottom) are also shown in dark cyan. For both plots, efficiency is found not to depend on the type of collisions, and, by comparing results, no loss in efficiency for beam-beam physics when injecting the gas is observed.}
	\label{fig:RTA_effs}
\end{figure}
\begin{itemize}
	\item \hltone, where particles are reconstructed in the tracking system and basic particle identification information from the calorimeter and the muon system is added. Selection algorithms, mostly inclusive with respect to several decay channels interesting to the \lhcb experiment, define which events are persisted to the disk buffer. The \hltone trigger level completely runs on GPUs~\cite{HLT1_CPU_GPU, Allen}, a major breakthrough for high-energy physics experiments of the \lhcb scale;
	\item \hlttwo, where offline-quality data reconstruction and selections are performed, including high-level particle identification from the \rich system. Selection algorithms optimized for specific hadron decays fulfilling the \lhcb physics program are then run.  
\end{itemize} 
Owing to the flexibility of the data acquisition system, concurrent reconstruction of beam-beam and beam-gas collisions was successfully achieved~\cite{CERN-THESIS-2021-313} by tuning the reconstruction algorithms to cope with the different collision geometries and energies. As an example, Fig.~\ref{fig:RTA_effs} shows the reconstruction efficiency of particles (top) and collision vertices (bottom) as a function of the collision longitudinal coordinate, by using the simulated samples introduced in Sec.~\ref{sec:simulations}.  Only the particles that cross the full tracker, reconstructed as \textit{long} tracks, and with minimum 3\gev momentum and 0.5\gev transverse momentum are considered. Two distinct peaks, corresponding to beam-gas and beam-beam collisions, are found in the distribution of the reconstructible particles (top) or PVs (bottom), being these defined as the particles leaving enough hits in the tracker system or the collisions producing at least three charged particles. The performance is found to be comparable for the two collision types and, by comparing the results for the different simulations, no loss in beam-beam efficiency because of the gas injection is found. Rather, because of the large difference in detector multiplicity between \pp and beam-gas collisions, a small decrease in the beam-gas tracking efficiency can be seen between the standalone and the overlapped data-taking scenarios. The interference between the two data types is proven to be minimal. Reconstructed events are then filtered according to a set of selection algorithms developed according to the physics case discussed in Sec.~\ref{sec:physics}. In particular, in order to reduce contamination from concurrent beam-beam collisions, a reconstructed collision vertex in the \textit{z} region covered by the SMOG2 cell is always required. Additionally, to suppress contamination between particles from \pp or proton-gas collisions in the same event, particles that are triggered on are required to originate or decay in the SMOG2 cell region.

\subsection{Luminosity}
\label{sec:Lumi}
As introduced in Sec.~\ref{sec:sc_principle}, within the storage cell, the gas density assumes a triangular profile with maximum value
\begin{equation}
    \rho_0=\frac{\Phi}{C_{tot}},
    \label{rho}
\end{equation}
where $\Phi$ is the gas particle rate (particle/s) and $C_{tot}$ the total conductance of the cell from the center outwards, also taking into account the gas temperature recorded by the dedicated probes.

\begin{figure}
    \centering
    \includegraphics [width=0.8\linewidth] {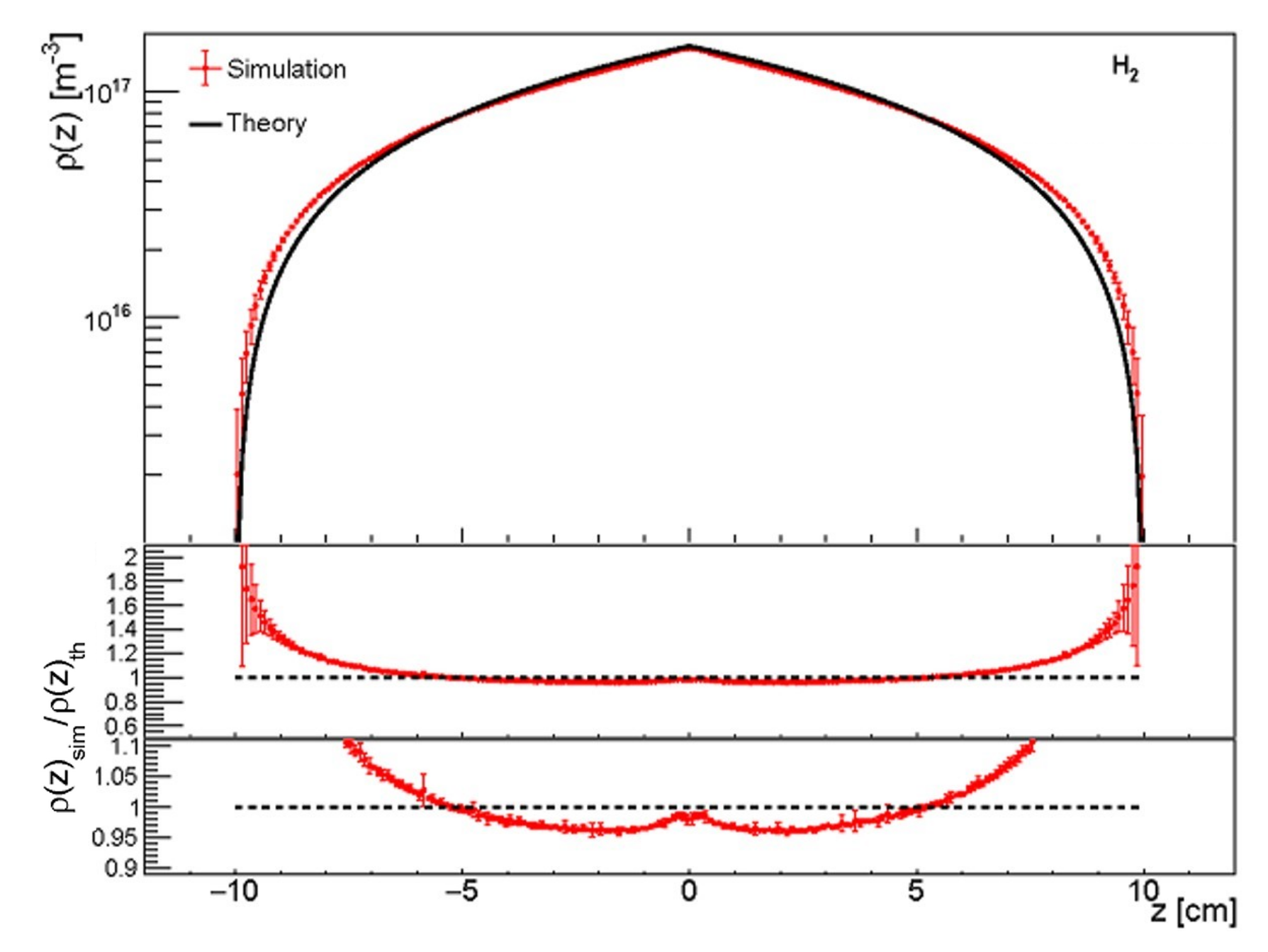}
    \includegraphics [width=0.8\linewidth] {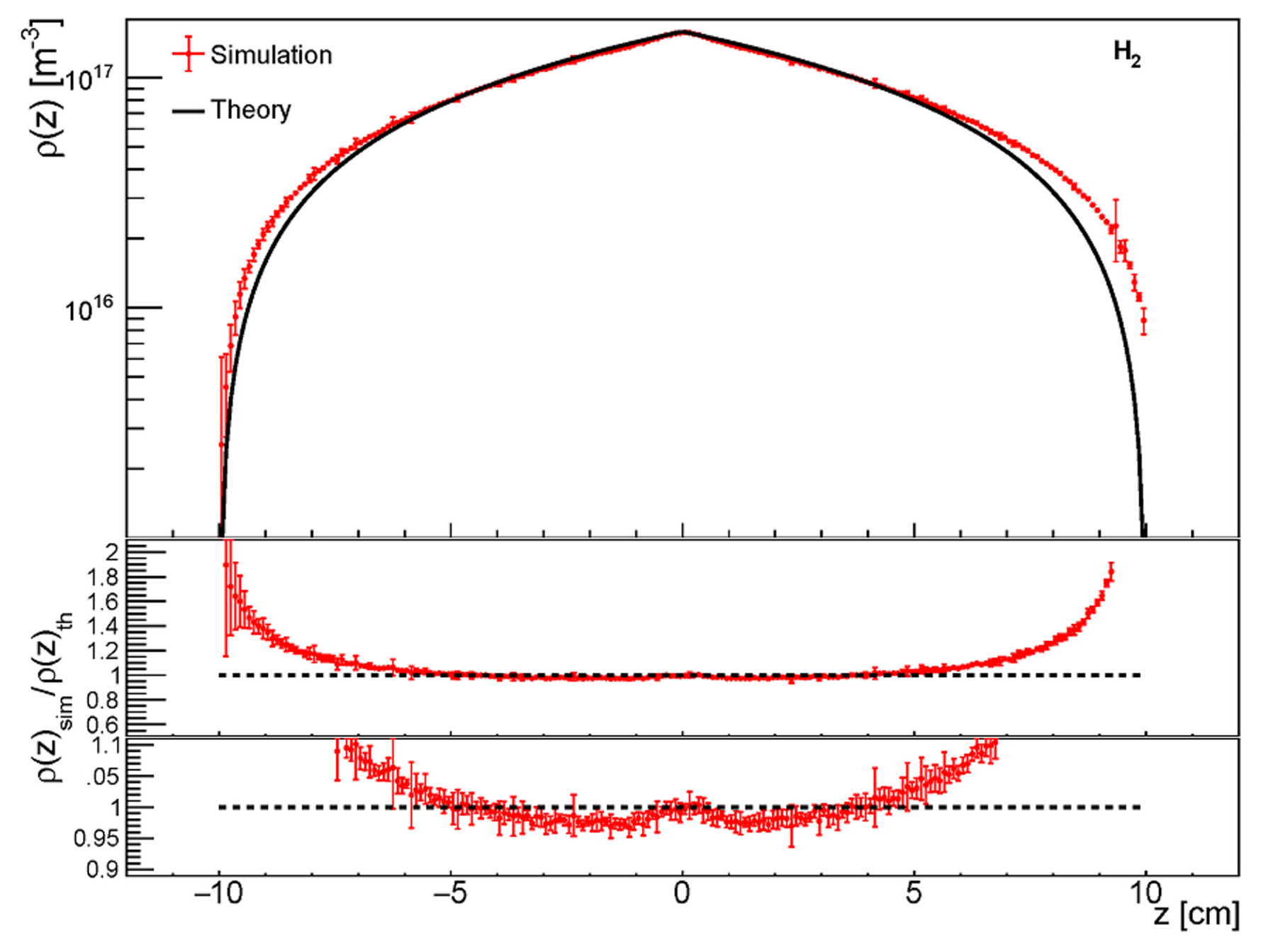}
    \caption{Gas density profile for H$_2$ for the cell only (top) and the cell with the RF-foils (bottom). The black line is obtained with the analytic method, the red dots with the \textit{Molflow+} simulation. The smaller panels show the ratio of the two curves, in the full range (middle) and in the SMOG2 cell central region (bottom) where the injected gas pressure is the highest.}
    \label{fig:Density-1}
\end{figure}

The average areal density values $\theta = \rho_0 \cdot L/2$, with L cell length,  are obtained by integrating the density profile, simulated by \textit{Molflow+}, along the beam axis.
A comparison between the simulation (Simu) and the analytical method (AM) is illustrated in Fig.~\ref {fig:Density-1} (top) for H$_2$. A very good agreement, differing the integrated areal density ratio $k = {\theta_{Simu}}/{\theta_{AM}}$ from one by 0.1\%, is found.

In the real configuration, the RF-foil effect also has to be taken into account. This affects the system conductance and limits the validity of the adopted approach assuming the ideal and isolated cell. A proper correction factor has hence been calculated and applied. Figure~\ref {fig:Density-1} (bottom) shows the density profiles in this configuration and  a clear deviation of the simulation from the theoretical expectation can be observed when approaching the right extreme of the cell. This difference has been estimated to be $k\sim$2.5\%, independent of the gas types and for gas fluxes ranging between \mbox{2 and 10 $\times \, 10^{-5}$ \,mbar\,$\cdot$l/s}, typical values of the injected gas flow. 

The luminosity is then given by
\begin{equation}
    \mathcal{L} = k \cdot \theta N_p f_{rev},
    \label{lumi}
\end{equation}
where $N_p = n_{bunch} \cdot n_{p/bunch}$ is the number of protons in the beam, given by the number of bunches ($n_{bunch}$) and number of protons per bunch ($n_{p/bunch}$), and $f_{rev} = 11\, 245 ~\rm{Hz}$ is the revolution frequency of the \lhc beams.
The systematic uncertainties related to the luminosity measurement have been estimated. These are due to the precision with which the parameters involved in the areal density estimation are known, such as the cell length and diameter, the gas temperature, and the injected gas flow. All the contributions remain relatively small, with the dominant one being the GFS accuracy, around 1\% if the gas flux is kept constant. The same calculations done for H$_2$ have been repeated for Ar and no difference has been found, as expected. In conclusion, the total systematic uncertainty on luminosity is expected as about 1.4\%, both for light and heavy gases.

\section{First results with 2022 LHC collisions}
%\section{output}
\label{sec:output}

\begin{figure*}
	\centering
	\includegraphics[width = 0.49\linewidth]{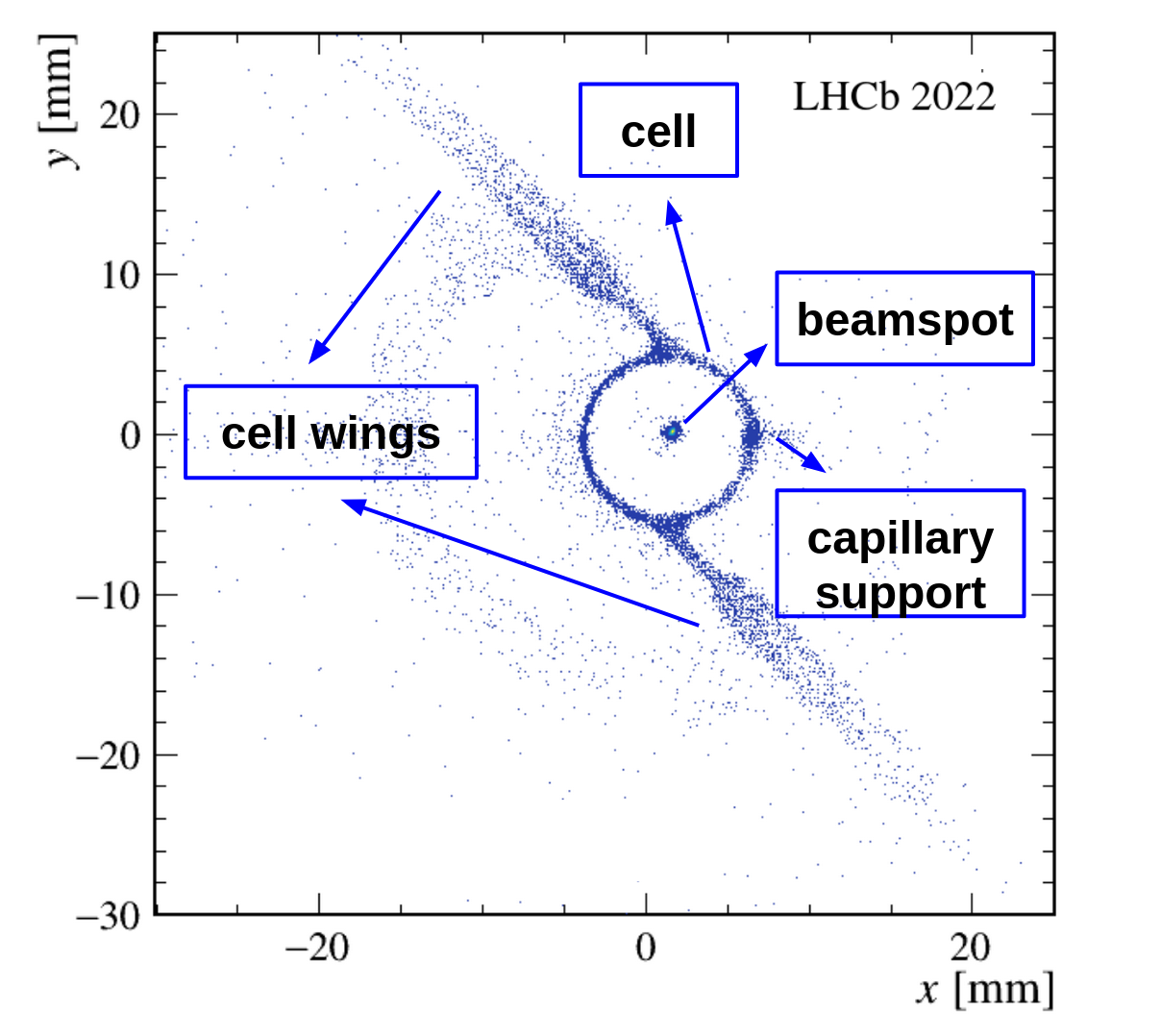}
 	\includegraphics[width = 0.49\linewidth]{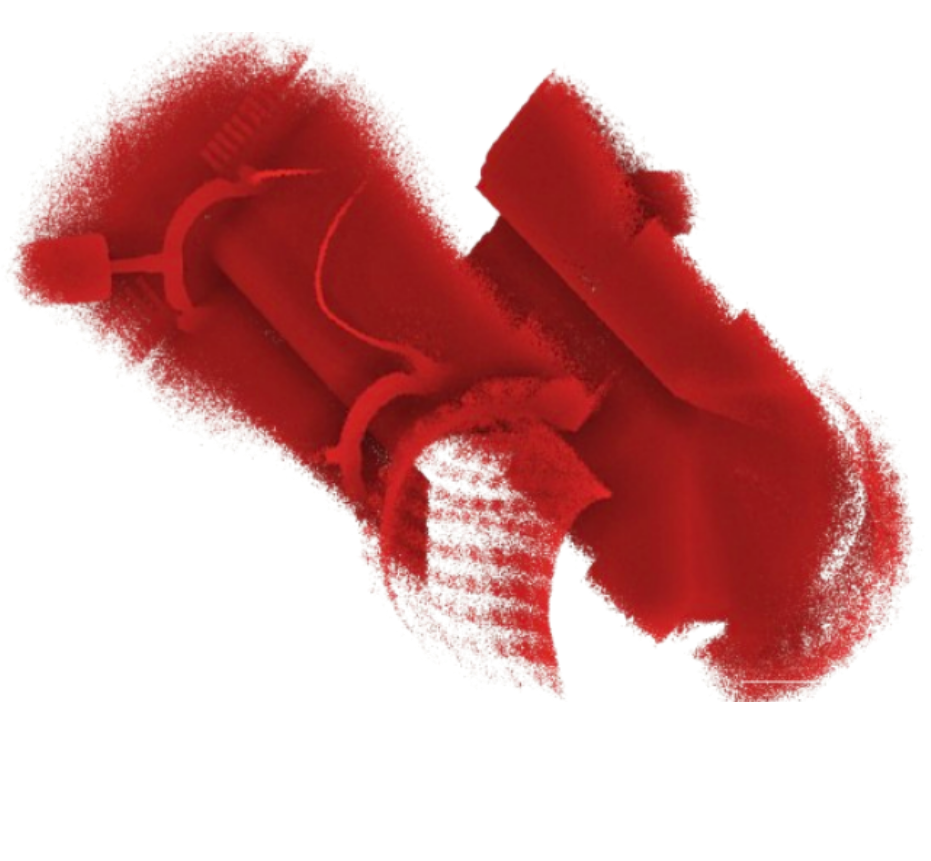}
	\caption{Tomography of the SMOG2 cell in its 2d closed (left) and its 3d open (right) positions as obtained by the reconstruction of material interaction vertices in 2022 and 2023 \pp collision data, respectively. By comparing e.g. with Fig.~\ref{fig:cell1a} and Fig.~\ref{fig:pedro1}, several elements of the SMOG2 hardware can be recognized.}
	\label{fig:SMOG2_cell_tomography}
\end{figure*}
As part of the 2022 \lhcb commissioning, data samples with injected helium, argon, neon, and hydrogen were collected. From the operational point of view, neither beam instabilities nor additional background have been observed due to the gas injections. Temperatures around $15 \, ^\circ$C were measured with no beam circulation, increased by about $25 \,^\circ$C during beam injection and tuning. The recorded maximum value is of $42 \,^\circ$C, well below the maximum tested one of 130$\,^\circ$C. The GFS was operated for all gases, giving very stable and precisely measured injected fluxes, as exemplified in Fig.~\ref{fig:GFS7}.

The collected data have then been analyzed. Firstly, reconstructed material interactions are employed to perform a tomography of the cell~\cite{VELO_mapping}, with the result shown in Fig.~\ref{fig:SMOG2_cell_tomography}. By exploiting 2022 collision data, the left bidimensional picture of the closed cell, of the two wings and of the support of the capillary where the gas flows through can be seen. Later, with 2023 data, the tridimensional representation was also obtained. By comparing e.g with the drawings in Fig.~\ref{fig:cell1a} or with the pictures in Fig.~\ref{fig:pedro1}, several elements can be recognised. By exploiting these figures, the aperture and the position of the cell relative to the \velo modules were measured with data, with all results consistent with expectations.

\begin{figure}
	\centering
	\includegraphics[width = \linewidth]{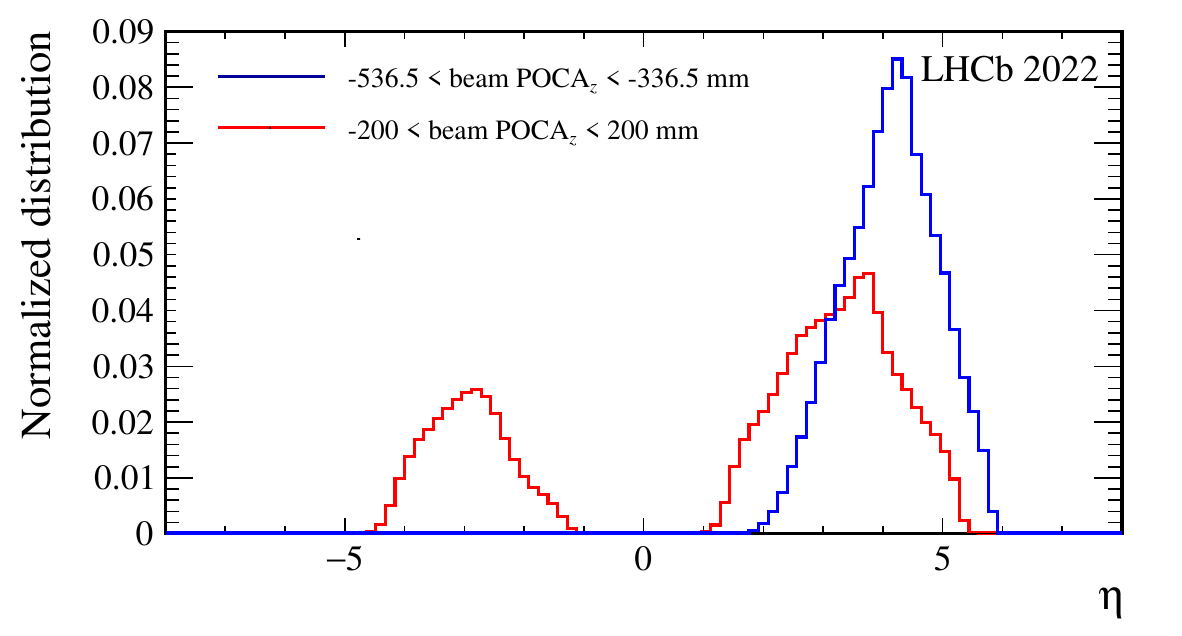}
	\includegraphics[width = \linewidth]{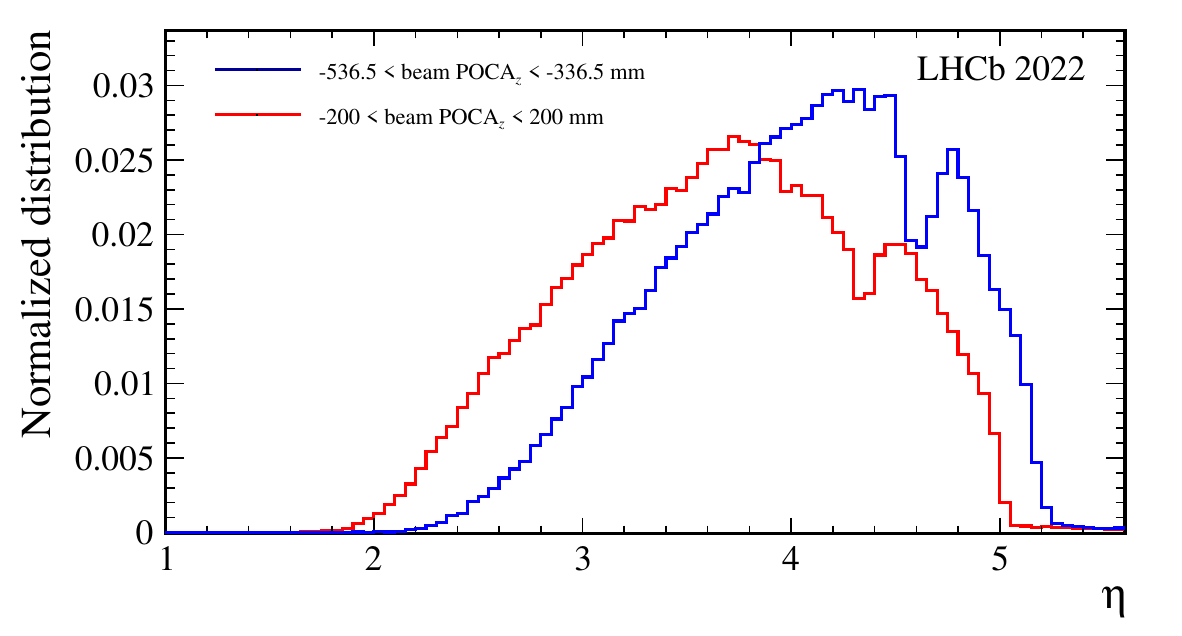}
	\caption{Normalized pseudorapidity distributions for particles reconstructed by the first trigger level in the \velo detector (top) and by the full tracking system (bottom) originating from \pAr (red) and \pp collisions (blue), as distinguished by the \textit{z} coordinate of their positions of closest approach (POCA$_{\textit{z}}$) to the beam axis.}
	\label{fig:SMOG2_tracking}
\end{figure}

The performance for reconstructing beam-beam and beam-gas collisions was then compared. Figure~\ref{fig:SMOG2_tracking} presents the pseudorapidity distributions\footnote{A particle pseudorapidity is defined as $\eta = -\ln[\tan(\theta/2)]$, being $\theta$ its polar angle with respect to the beam direction, conventionally assumed as the $z$ one.} for the particles reconstructed in the first trigger level. Particles produced in \pp collisions (in red) are symmetric in the \velo detector (top), modulo the inefficiency coming from a lower number of \velo modules in the backward region. Only forward particles, instead, originate from the fixed-target collision (in blue), as expected. Such a difference is then reduced for particles reconstructed by the full tracking system (bottom), where the spectrometer acceptance is for particles produced at the nominal interaction point with $\eta \in [2, 5]$. 
\begin{figure}
	\centering
	\includegraphics[width = \linewidth]{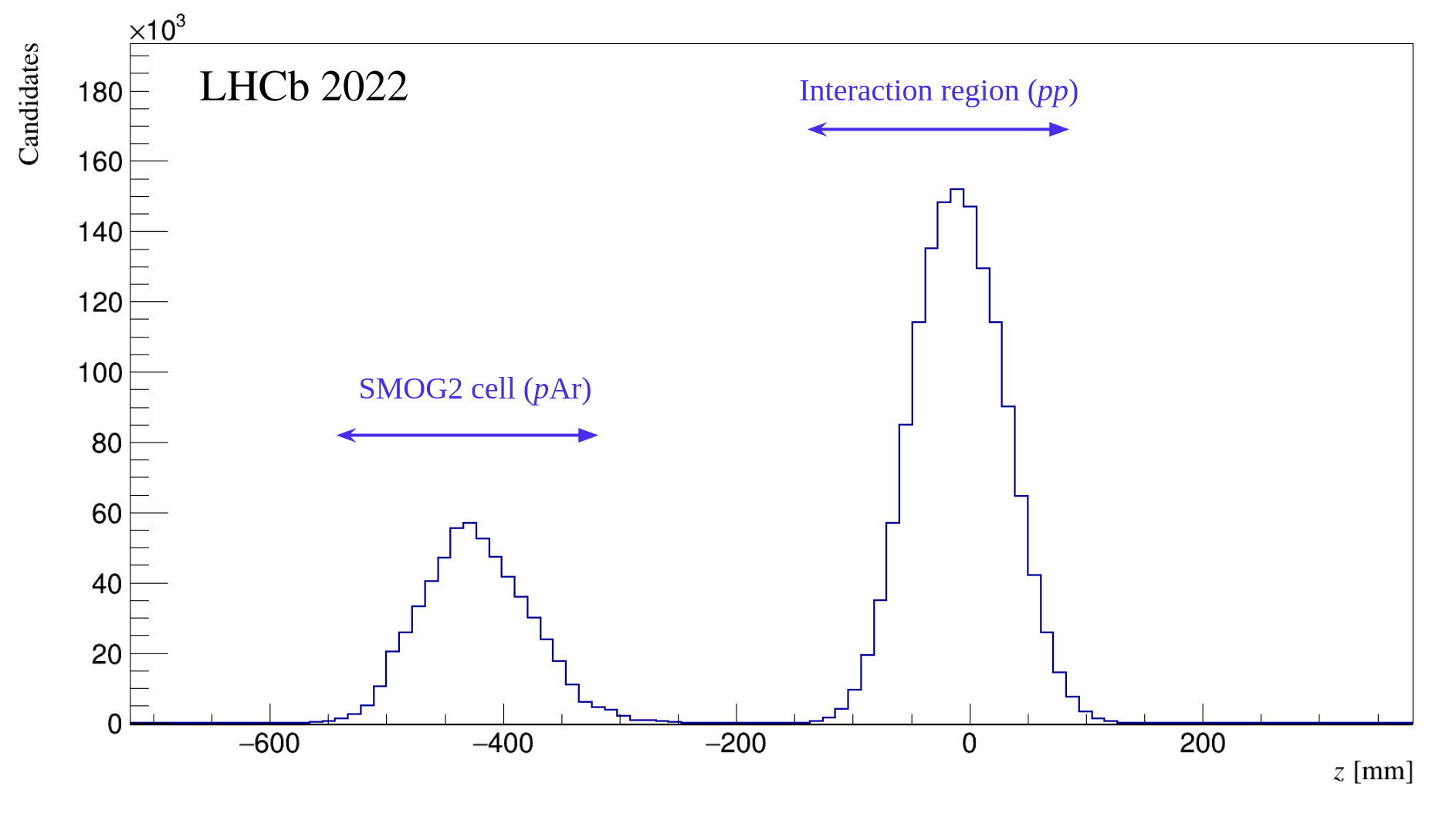}
	\caption{Distributions of the \textit{z} coordinate for collision vertices reconstructed in the first trigger level during a data-taking with overlapped \pp and \pAr collisions. Two distinct distributions, following a Gaussian and a triangular-like  distributions, respectively, can be clearly seen.}
	\label{fig:SMOG2_PVs_1}
\end{figure}

\begin{figure}
	\centering
	\includegraphics[width = \linewidth]{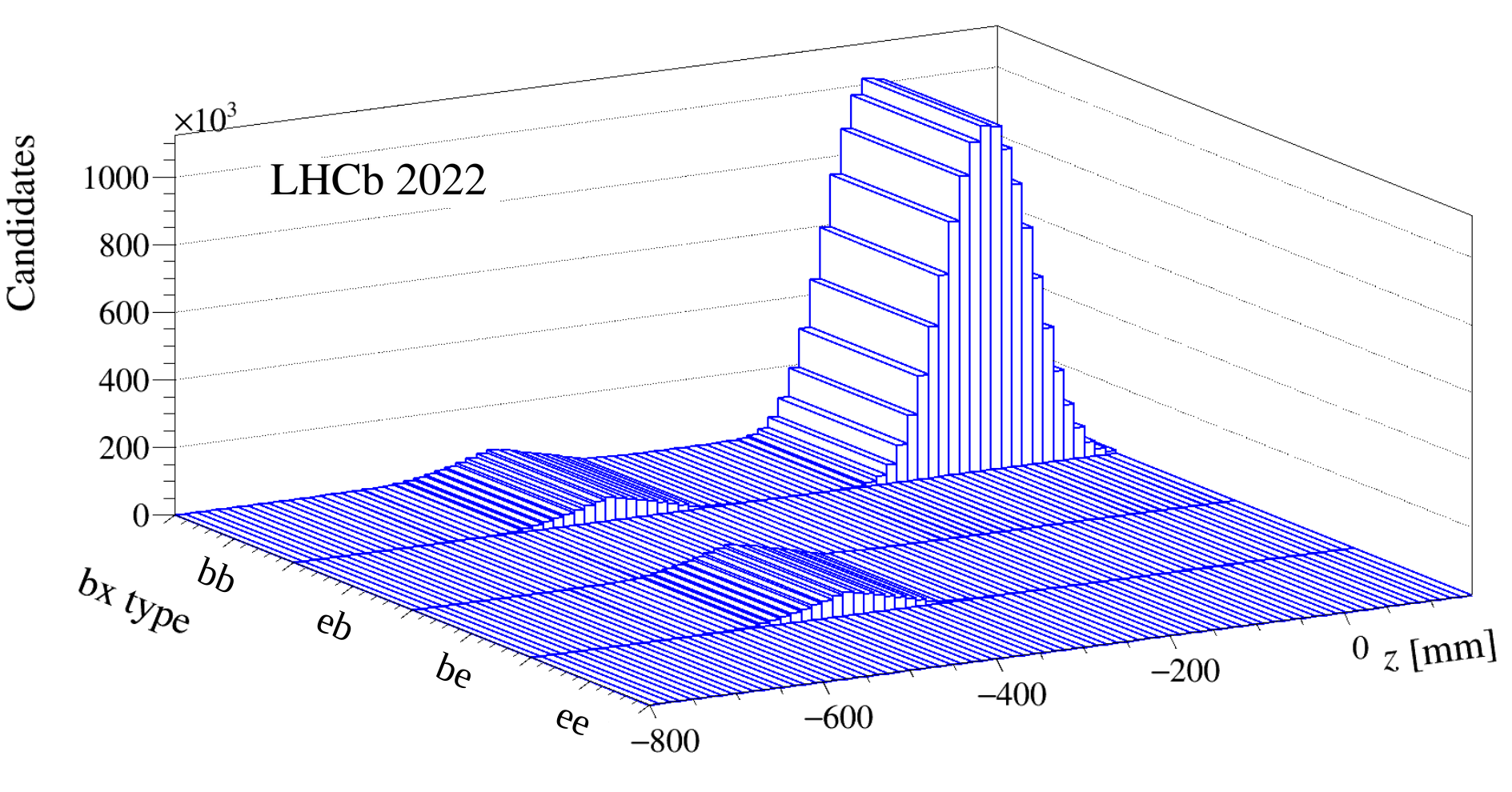}
	\includegraphics[width = \linewidth]{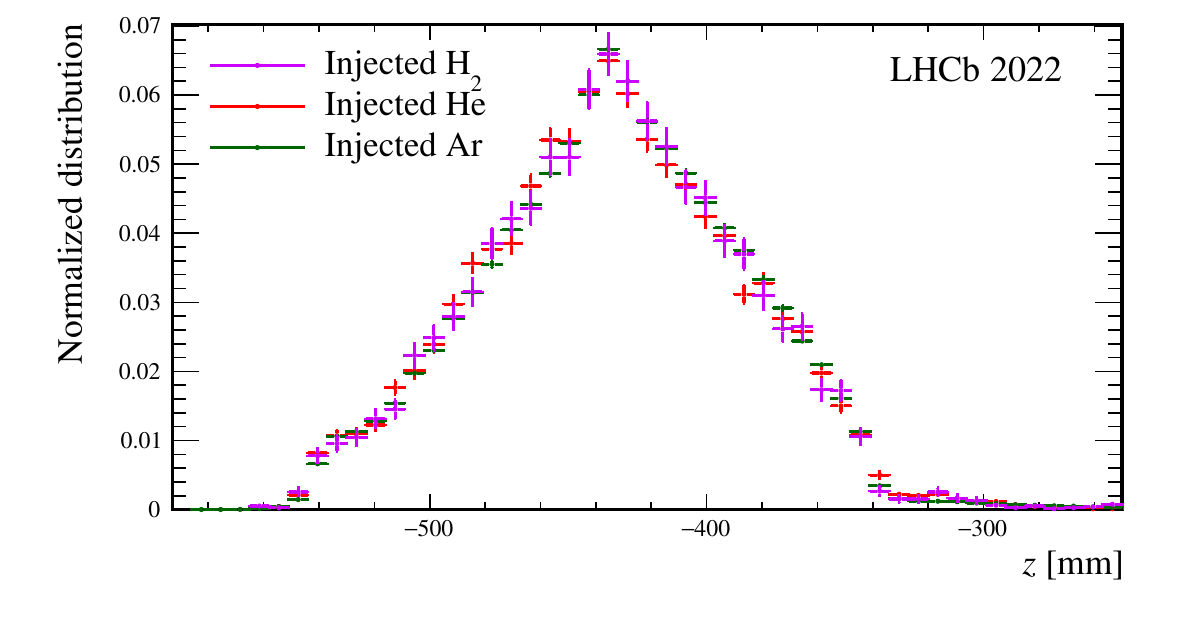}
	\caption{Distributions for collision vertices reconstructed in the first trigger level. The top plot shows, in a data-taking with injected helium, the reconstructed vertices for different \lhc bunch crossing configuration. The bottom plot compares the PV$_z$ distributions in the SMOG2 cell with different injected gas and they are found to be compatible.}
	\label{fig:SMOG2_PVs_2}
\end{figure}

\begin{figure}
\centering
\includegraphics[width=\linewidth]{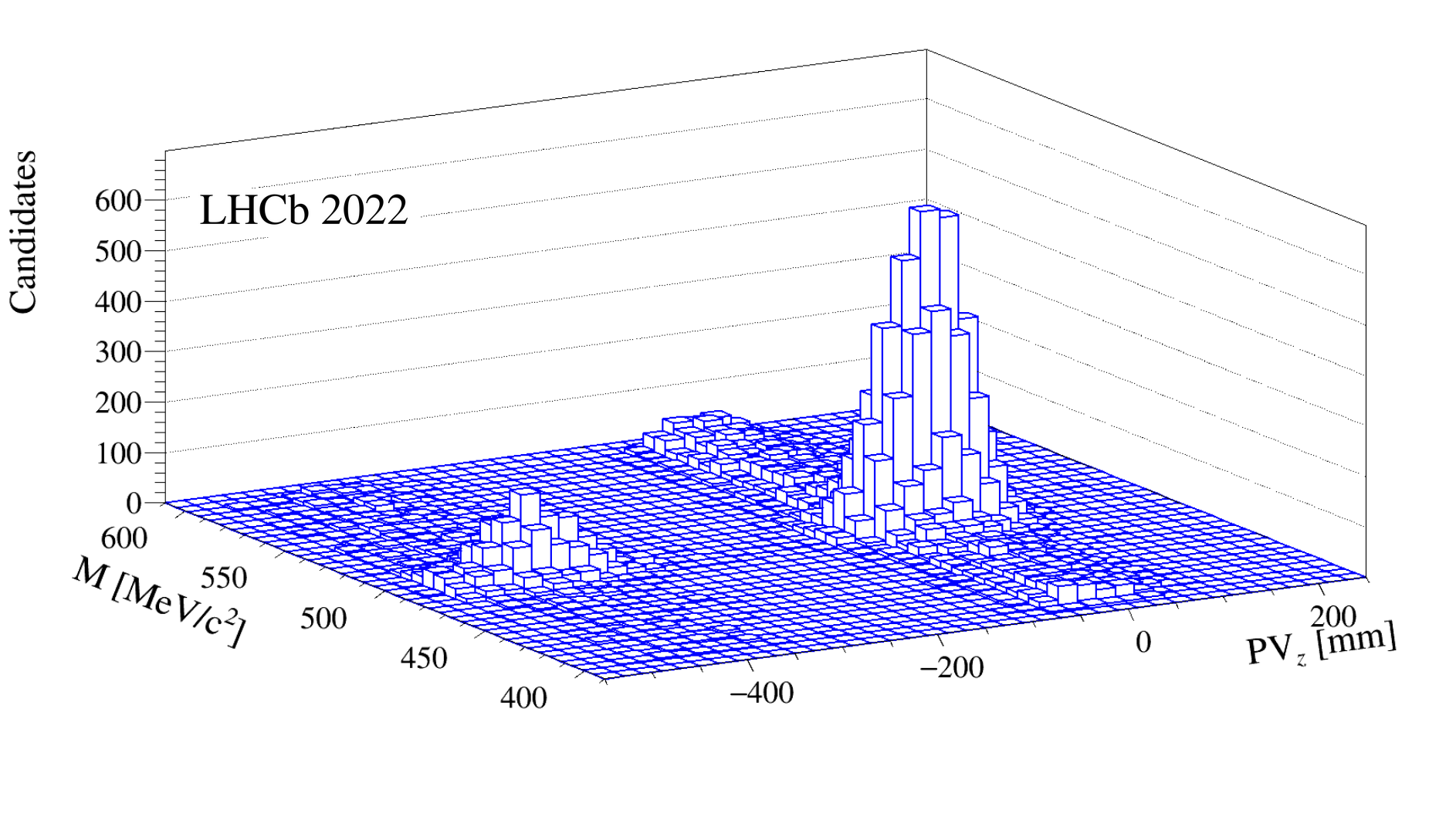}
\includegraphics[width=\linewidth]{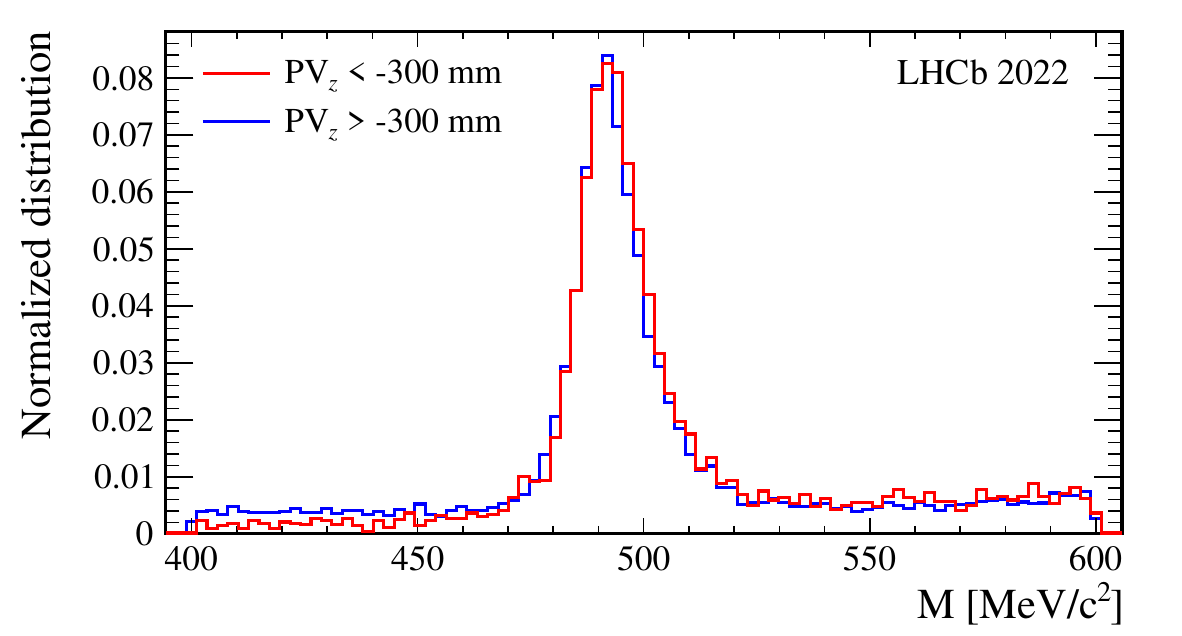}
\caption{Invariant mass distributions for $\decay{\KS}{\pip\pim}$ decays reconstructed in the first trigger level from \pppAr collisions. Top plot shows the 
distributions as a function of invariant mass and PV$_z$ associated to the \KS particle and two clear peaks, corresponding to \pAr and \pp collisions, emerge. The \KS invariant mass for \pAr (red) and \pp (blue), compared in the bottom plot, have compatible widths.}
\label{fig:SMOG2_KS}
\end{figure}
Collision vertex reconstruction is then illustrated in Figs.~\ref{fig:SMOG2_PVs_1} and \ref{fig:SMOG2_PVs_2}. In the former, the distribution for the longitudinal coordinate of \pp and \pAr vertices reconstructed in the first trigger level, is shown. Evidently, the \lhcb detector, when injecting gas, can be claimed as equipped with two distinct and independent collision points, allowing to study collisions in two different systems and with two energy scales. In top Fig.~\ref{fig:SMOG2_PVs_2}, the \textit{z} vertices distribution is illustrated for different \lhc bunch-crossing types\footnote{Conventionally, the \lhc bunch crossing configuration is expressed as two letters, one for the clockwise, beam1, and one for the counter-clockwise, beam2, \lhc beam. A b for beam1 or beam2 means the bunch is filled with particles, an e that the bunch is empty.}. Independently on the \pp collisions happening in \lhcb, all \lhc beam1 bunches, entering \lhcb from the \velo, can be exploited to collect fixed-target data. The bottom plot compares the \textit{z} vertex distributions from runs with different injected gases. By comparing with the expected profiles obtained with the \textit{Molflow+} simulations discussed in Sec.~\ref{sec:Lumi}, a qualitative agreement can also be seen. 

Particles reconstructed in the full spectrometer are then paired to form secondary vertices where the decay of composite particles occurs. As an example, invariant mass distribution reconstructed in the first trigger level for \decay{\KS}{\pip\pim} candidates is presented in Fig.~\ref{fig:SMOG2_KS}. As a function of the collision vertex \textit{z} coordinate associated with the \KS meson (top plot), two distinct peaks, corresponding to \pp and \pAr collisions, clearly appear. Moreover, by projecting them on the invariant mass axis (bottom), the widths of the mass peak are found to be comparable, demonstrating the detector momentum resolution is only slightly \textit{z}-dependent. The coarse consistency between the peak center values also proves good control on the momentum scaling, modulo minimal effects due to the non perfect detector calibration in the 2022 commissioning phase. 

\begin{figure}
\centering
\includegraphics[width=\linewidth]{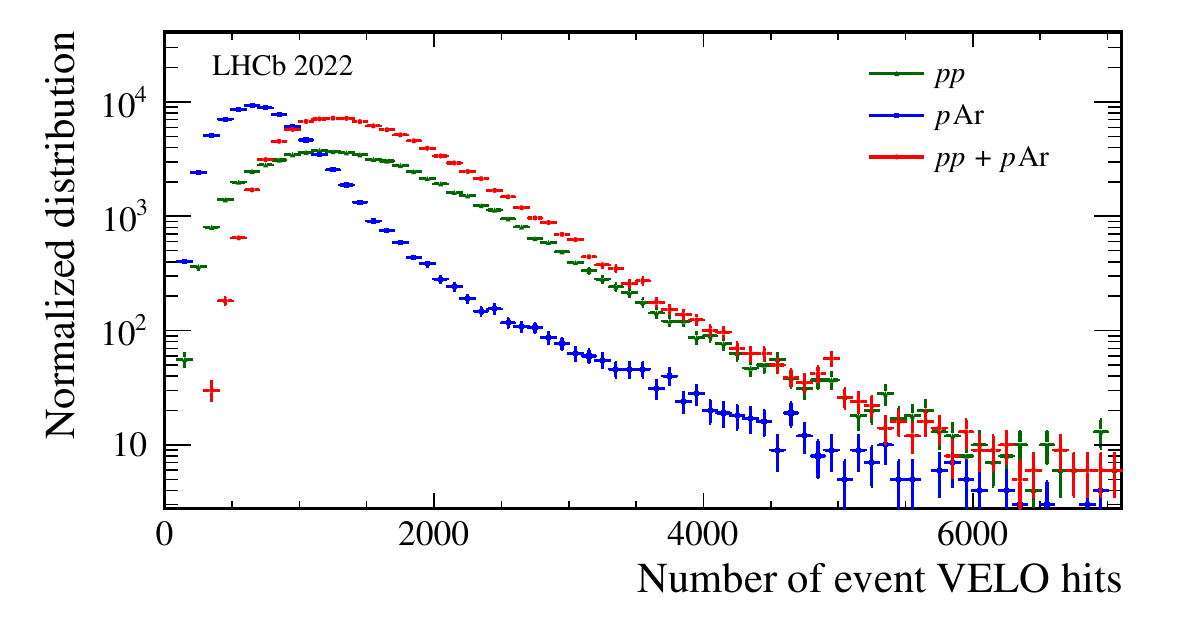}
\includegraphics[width=\linewidth]{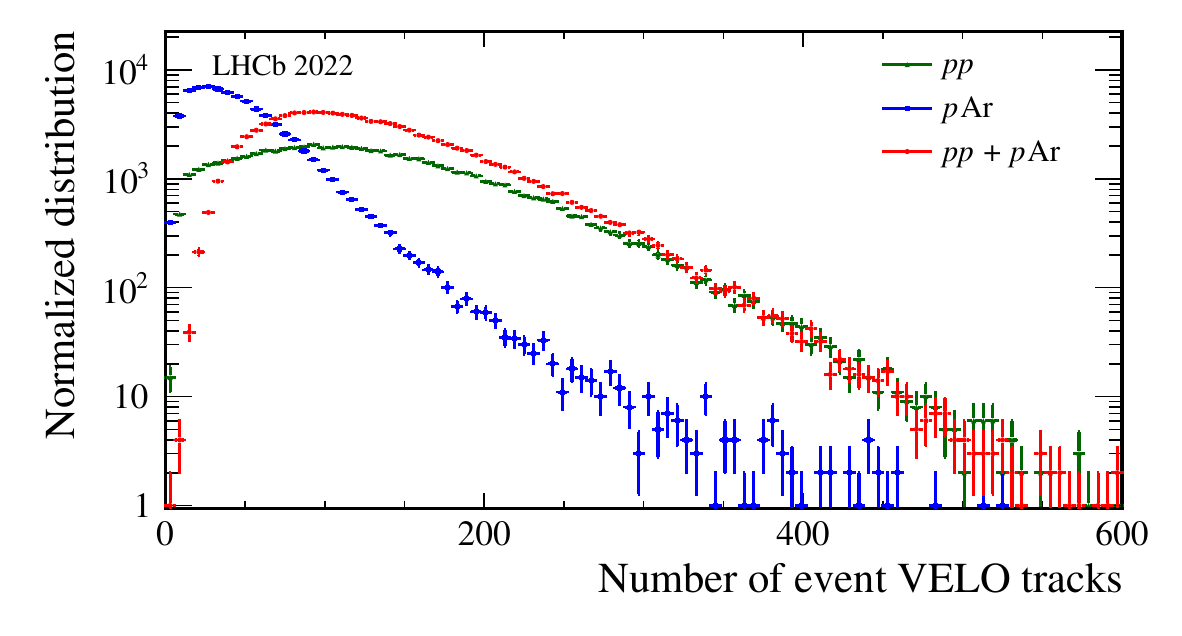}
\caption{Normalized distributions of the number of hits (top) and tracks (bottom) reconstructed in the first trigger level in bunch crossings with reconstructed collision vertices only in the SMOG2 cell (blue), in the \lhcb \pp interaction point (in green) or in both (in red). For both figures, the averaged number of per-bunch \pp collision was about one.}
\label{fig:VELO_comparisons}
\end{figure}

While the presented distributions clearly demonstrate that collisions in the SMOG2 cell can be reconstructed, the negligible interference with the \pp data-taking had to be demonstrated as well. In Fig.~\ref{fig:VELO_comparisons}, the number of \velo hits (top) and tracks (bottom) for collisions with reconstructed vertices only in the SMOG2 area (blue), only in the \pp interaction point (green) and in both (red) are compared. The increase in the \velo occupancy, whose reconstruction dominates the data processing time, is small when injecting the gas (4.2\% for the number of \velo hits and 1.5\% for the number of \velo tracks). On top of this, it is worth reminding that, while during the 2022 commissioning only about one per-bunch \pp collisions was used, the nominal Run 3 value will be larger. The effect of gas injection on the \pp data-taking will decrease correspondingly.

\begin{figure}
\centering 
\includegraphics[width = \linewidth]{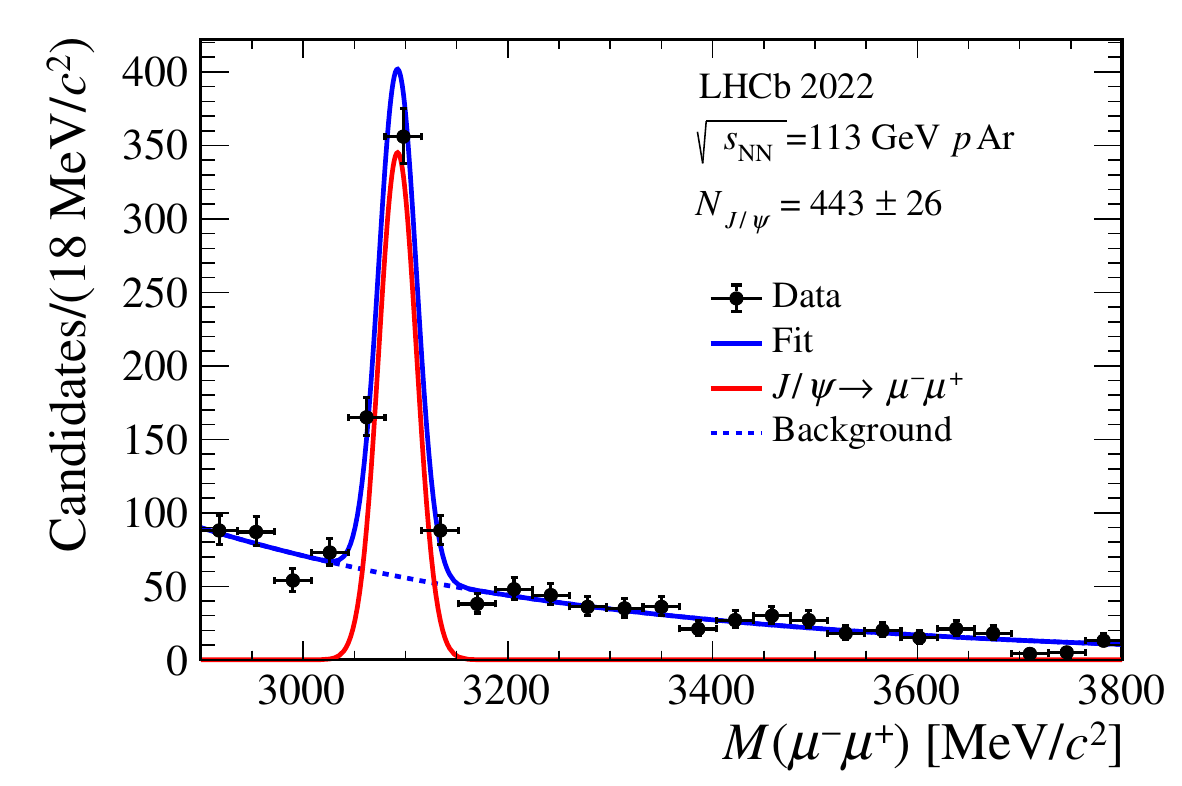}
\includegraphics[width = \linewidth]{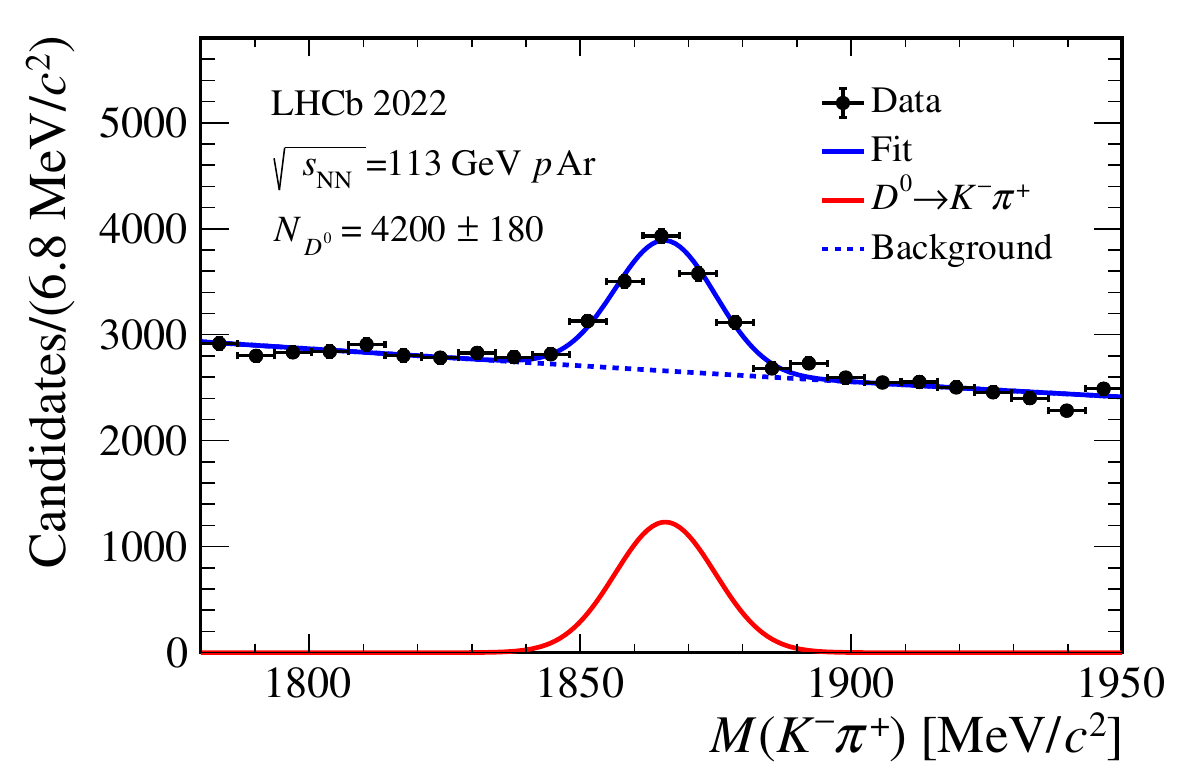}
\caption{Invariant mass distributions for (top) \jpsi and (bottom) \Dz decays as resulting from the full data processing chain in 18 minutes of data-taking with injected argon. Both mass peaks are modeled with a Gaussian function for the signal and an exponential for the background.}
\label{fig:offline_mass_peaks-1}
\end{figure}

\begin{figure}
\centering 
\includegraphics[width = \linewidth]{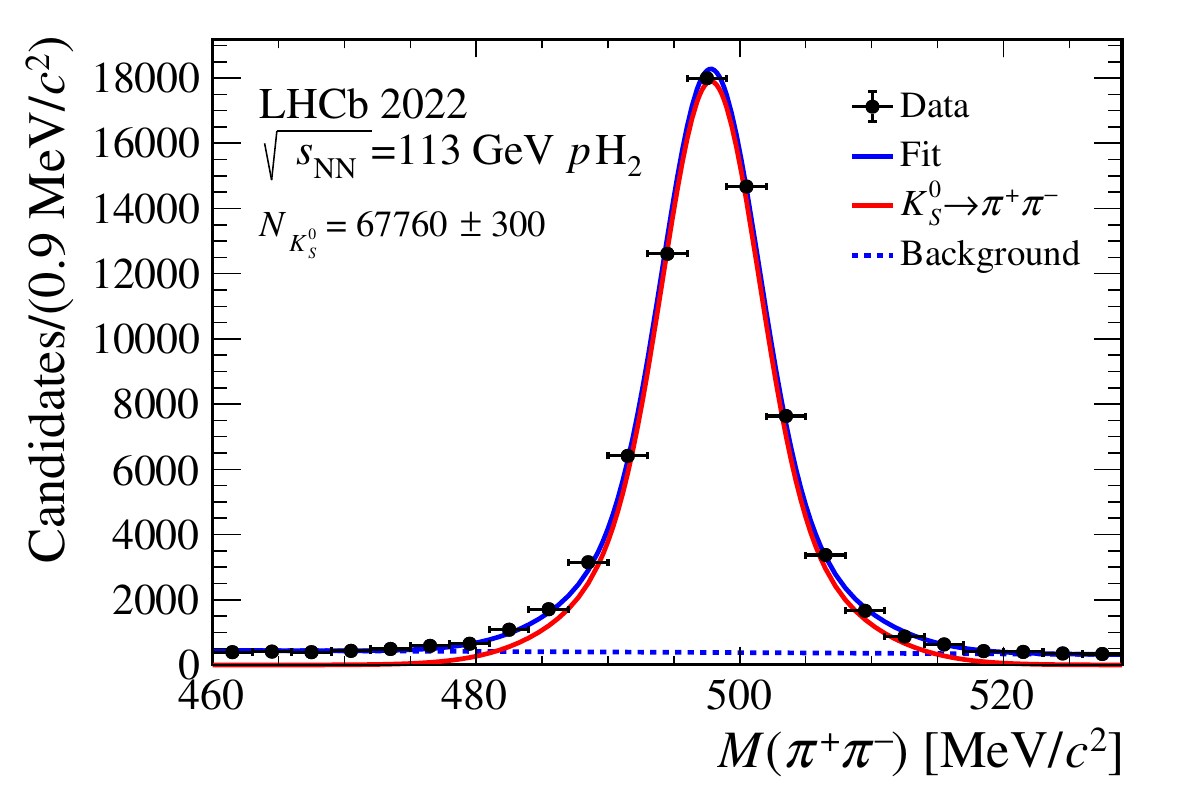}
\includegraphics[width = \linewidth]{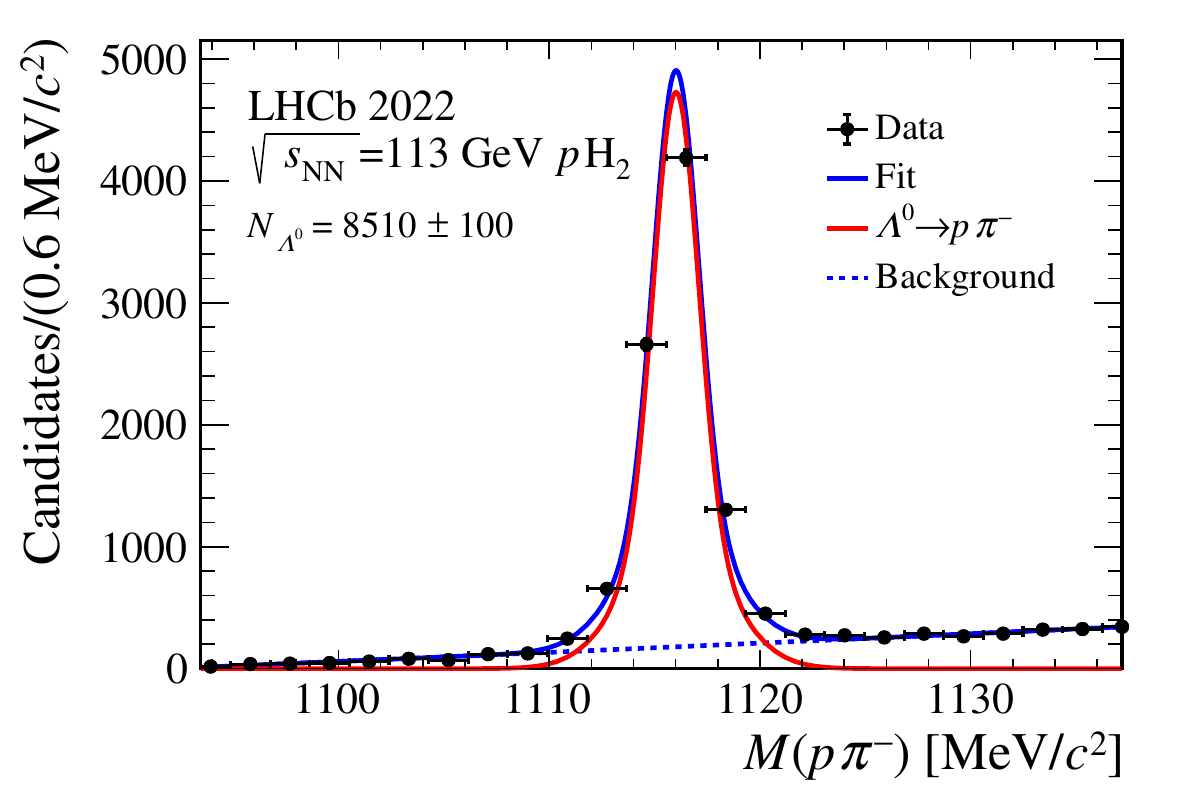}
\caption{Invariant mass distributions for (top) \Lz and (bottom) \KS decays as resulting from the full data processing chain in 20 minutes of data-taking with injected hydrogen. Both mass peaks are modeled with a Gaussian function for the signal and an exponential for the background.}
\label{fig:offline_mass_peaks-2}
\end{figure}

The validation of the full reconstruction chain, namely the two software trigger levels and the offline analysis infrastructure, has also been performed for SMOG2 events. Examples of performance results are presented in Figs.~\ref{fig:offline_mass_peaks-1} and \ref{fig:offline_mass_peaks-2}. In Fig.~\ref{fig:offline_mass_peaks-1}, a data sample with injected argon in the SMOG2 cell is considered, and \jpsi (top) and \Dz (bottom) candidates are reconstructed and selected. For both composite particles, the resolution is found to be comparable with similar analyses of \pp collisions data. Plots of Fig.~\ref{fig:offline_mass_peaks-2} show, on a data sample with injected hydrogen in the SMOG2 cell, the \KS (top) and \Lz (bottom) candidates invariant mass distribution. It is worth to underline that these figures result from a data-taking of just 18 and 20 minutes, respectively. As discussed in Sec.~\ref{sec:physics}, high statistics and efficiently reconstructed charm channels produced in fixed-target collisions will be available with SMOG2.

\section{Conclusions}
%\section{Conclusions}

\label{sec:Conclusions}

The upgrade of the \lhcb gas target was possible 
through a collaborative R\&D effort involving \lhcb and \lhc teams. For the first time, a storage cell able to deliver a large amount of beam-gas collisions without perturbing the beam-beam collision system or the beam lifetime was installed in the \lhc primary vacuum. The system has been extensively commissioned in 2022 with several injections, notably of a non-noble gas for the first time. No beam instability has been reported, validating the studies discussed throughout the paper and excluding detrimental interactions of the SMOG2 cell with the \lhc machine. No unexpected response by the temperature probes has been observed, excluding effects of heating from the beam. The GFS was operated injecting all gases, with excellent stability and precise reproducibility of the injected fluxes, which is of paramount importance to have a reliable luminosity measurement in real time. Physics channels have been studied with the collected beam-gas collisions. With injections lasting only about 20 minutes, large samples of charm hadron signals are reconstructed, with momentum resolution and efficiencies that are mostly comparable between beam-beam and beam-gas collisions. Finally, only small increases in the detector multiplicity have been observed because of the beam-gas collisions. Overall, \lhcb is demonstrated to be capable of efficiently managing and processing data from both collision systems without compromising its performance.

The implementation of SMOG2 is expected to enhance the fixed-target program in several ways. These improvements include expanding the range of available gas species for experimentation, enabling better control over target gas pressure and instantaneous luminosity, and significantly increasing the integrated luminosities of the fixed-target samples by up to two orders of magnitude. Furthermore, the use of hydrogen and deuterium in particular will serve as a reference for measurements involving heavier nuclei. It will also enable measurements of nucleon structure in a novel kinematic regime, providing valuable insights into the properties and behavior of nucleons in different experimental conditions. As well, events reproducing primary cosmic-ray collisions in the interstellar medium and in the atmosphere will provide insights to astroparticle physics. 
At the same time, heavier targets, such as Argon, Krypton or Xenon, can extend the studies of nuclear
matter in a domain where QGP effects are expected to be manifest. 

A possible future upgrade, known as LHCspin~\cite{LHCSpin_1,LHCSpin_2,LHCSpin_3,LHCSpin_4}, has already performed several R\&D studies and represents the natural evolution of SMOG2 aiming at installing
a polarized gas target opening the door to spin physics at \lhc. 
With strong interest and support from the international theoretical community, LHCspin could be
a unique opportunity to advance our knowledge on several unexplored QCD areas, complementing
both existing facilities and the future Electron-Ion Collider~\cite{EIC_1, EIC_2}.

\section*{Acknowledgements}
%\section{Acknowledgements}
\label{sec:acknow}

We are grateful to Ulrich Uwer for his critical reading and helpful comments. We thank the whole LHCb collaboration, especially the operation teams who put significant effort to collect data during the commissioning phase of the upgraded experiment. We also thank the technical and administrative staff at the LHCb institutes. We acknowledge support from CERN and from the national agencies: 
CNRS/IN2P3 (France); FAU (Germany); INFN (Italy); NWO (Netherlands).
Support from the European Union's Horizon 2020 research and innovation program (grant agreement No.\ 824093, STRONG-2020, Working Package ``JRA2-FTE@LHC") is acknowledged.

\setboolean{inbibliography}{true}
\bibliographystyle{LHCb}
\addcontentsline{toc}{section}{References}
\bibliography{main,gfs,sc,LHCb-TDR,LHCb-DP,LHCb-PAPER,physics}
%\bibliography{Operation,main,LHCb-PAPER,LHCb-CONF,LHCb-DP,,report,physicsref,standard,rrb}
\end{document}